\documentclass[preprints,article,accept,moreauthors,pdftex]{Definitions/mdpi}
\usepackage[labelformat=simple]{subfig}

\firstpage{1} 
\pubvolume{1}
\issuenum{1}
\articlenumber{0}
\pubyear{2021}
\copyrightyear{2021}
\datereceived{} 
\dateaccepted{} 
\datepublished{} 
\hreflink{https://doi.org/} % If needed use \linebreak
\Title{The I-Love-Q Relations for Superfluid Neutron Stars}
\TitleCitation{The I-Love-Q Relations for Superfluid Neutron Stars}
\Author{Cheung-Hei Yeung $^{1}$, Lap-Ming Lin $^{1,}$*, Nils Andersson $^{2}$ and Greg Comer $^{3}$}

\AuthorNames{Cheung-Hei Yeung, Lap-Ming Lin, Nils Andersson and Greg Comer}
\AuthorCitation{Yeung, C.-H.; Lin, L.-M.; Andersson, N.; Comer, G.L.}
\address{%
$^{1}$ \quad Department of Physics, The Chinese University of Hong Kong, Hong Kong, China; chyeung@phy.cuhk.edu.hk \\
$^{2}$ \quad Mathematical Sciences and STAG Research Centre, University of Southampton, Southampton
SO17 1BJ, UK; n.a.andersson@soton.ac.uk
\\
$^{3}$ \quad Department of Physics, Saint Louis University, St. Louis, MO 63156-0907, USA; greg.comer@slu.edu}

\corres{Correspondence:lmlin@cuhk.edu.hk}

\abstract{The I-Love-Q relations are approximate equation-of-state independent relations that
connect the moment of inertia, the spin-induced quadrupole moment, and the tidal deformability of 
neutron stars. 
In this paper, we study the I-Love-Q relations for superfluid neutron stars for a general relativistic two-fluid model: one fluid being the neutron superfluid and the other a conglomerate 
of all charged components. 
We study to what extent the two-fluid dynamics might affect the robustness of the I-Love-Q
relations by using a simple two-component polytropic model and a relativistic mean field model 
with entrainment for the equation-of-state. 
Our results depend crucially on the spin ratio $\Omega_{\rm n}/\Omega_{\rm p}$ between the angular velocities of the neutron superfluid and the normal component. 
We find that the I-Love-Q relations can still be satisfied to high accuracy for superfluid 
neutron stars as long as the two fluids are nearly co-rotating 
$\Omega_{\rm n}/\Omega_{\rm p} \approx 1$. 
However, the deviations from the I-Love-Q relations increase as the spin ratio deviates 
from unity. In particular, the deviation of the Q-Love relation can be as large as $O(10\%)$ 
if $\Omega_{\rm n}/\Omega_{\rm p}$ differ from unity by a few tens of percent. 
As $\Omega_{\rm n}/\Omega_{\rm p} \approx 1$ is expected for realistic neutron
stars, our results suggest that the two-fluid dynamics should not affect the accuracy of 
any gravitational waveform models for neutron star binaries that employ the relation to connect 
the spin-induced quadrupole moment and the tidal deformability. }

  % List three to ten pertinent keywords specific to the article, yet reasonably common within the subject discipline.

\begin{document}
%%%%%%%%%%%%%%%%%%%%%%%%%%%%%%%%%%%%%%%%%%

%%%%%%%%%%%%%%%%%%%%%%%%%%%%%%%%%%%%%%%%%%
%\setcounter{section}{-1} %% Remove this when starting to work on the template.

\section{Introduction}
\label{sec:intro} 

The groundbreaking detection of the first gravitational wave signal from a binary neutron star 
system GW170817~\cite{GW170817} has opened up a powerful channel to study the internal 
structures of neutron stars and the poorly understood supranuclear equation-of-state (EOS).  
In particular, the~signal from the GW170817 event has been used to constrain 
the tidal deformability and the implications for EOS models have also been studied extensively
(see, e.g.,~\cite{Annala:2018,De:2018,Fattoyev:2018,Most:2018,Tews:2018,Lim:2018,Malik:2018,Li:2019,
Carson:2019}). 
Observations of further gravitational wave events involving binary systems, or even isolated neutron stars, with increasing precision will help us get closer to unlocking the secrets of high density nuclear~matter.

Observations of neutron stars can provide important information about the 
poorly understood nuclear matter EOS due to the fact that the properties of neutron stars, 
in general, depend sensitively on the matter model. It is thus quite surprising that various approximately EOS-insensitive relations connecting different neutron star properties have been discovered in the past decade~\cite{Tsui:2005,Lau:2010,Yagi:2013a,Yagi:2013b,Yagi:2014,Chan:2014, 
Chakrabarti:2014,Pappas:2014,Pappas:2015,Breu:2016,Bozzola:2017,Luk:2018,Riahi:2019,Sun:2020}. 
These relations are ``universal'' in the sense that they are insensitive to the EOS models to the $O(1\%)$ level (see~\cite{Yagi:2017,Doneva:2018} for reviews). 
These universal relations can be powerful tools to infer the physical quantities of neutron stars.
For instance, by~making use of the universal relation for the $f$-modes of neutron stars~\cite{Lau:2010}, the~mass and radius of an isolated neutron star can be inferred accurately if gravitational waves emitted from the $f$-mode oscillations of the star can be detected.

Universal relations can also help reduce the number of parameters in theoretical gravitational
waveform models for binary neutron star inspirals~\cite{Lackey:2019,Schmidt:2019,Andersson:2019,Barkett:2020}. 
It is noted that the I-Love-Q relations discovered by Yagi and Yunes~\cite{Yagi:2013a,Yagi:2013b} connect the moment of inertia $I$, the~$l=2$ tidal deformability, and~the spin-induced quadrupole moment $Q$ of slowly rotating neutron stars. The~multipole Love relation discovered by Yagi~\cite{Yagi:2014} connects the $l=2$ and $l=3$ tidal deformabilities. The~f-mode-Love relations found by Chan~et~al.~\cite{Chan:2014} relate the $f$-mode frequency to the tidal deformability. 
For waveform models that include the quadrupolar ($l=2$) and octopolar ($l=3$) adiabatic 
and dynamical tidal effects~\cite{Andersson:2019}, these universal relations~\cite{Yagi:2013a,Yagi:2013b,Yagi:2014,Chan:2014} can be used to reduce the intrinsic matter
parameters from ten to two, namely the $l=2$ (scaled) tidal deformabilities of the two stars. 
The other eight matter parameters include the spin-induced quadrupole moments, the~$l=3$ tidal 
deformabilities, and~the $l=2$ and $l=3$ $f$-mode frequencies of the two~stars. 

In view of their relevance for neutron-star astrophysics and gravitational-wave physics, it is 
important to test the robustness of these universal relations. 
Although they have been demonstrated to be insensitive to many EOS models to the $O(1\%)$ level, 
it is not yet well established whether these universal relations are also insensitive to 
more realistic neutron star physics aspects, like the state of matter at high densities.
Work in this direction includes the extension of the I-Q relation for rapidly rotating 
stars. While it was originally found that the relation becomes more EOS-dependent when considering 
rapidly rotating stars with a fixed rotation frequency~\cite{Doneva:2013}, it was later found that the I-Q relation, in fact, remains approximately EOS-insensitive if one uses dimensionless parameters to characterize the rotation~\cite{Pappas:2014,Chakrabarti:2014}. 
The I-Q relation is also found to fail for slowly rotating neutron stars with very strong 
magnetic fields so that the stellar deformation is dominated by the magnetic field instead of 
rotation~\cite{Haskell:2013}.  
Thermal effects relevant for protoneutron stars have also been shown to break the I-Love-Q relations~\cite{Martinon:2014,Marques:2017}. 
It has also been demonstrated that the I-Love-Q relations are still satisfied to within 
$\sim 3$\% for hybrid star models with a strong first-order hadron-quark phase transition in the interior~\cite{Paschalidis:2018}. 
However, the~I-Love relation can be broken if the quark matter core is in a crystalline 
color-superconducting state~\cite{Lau:2017,Lau:2019}. 
In this paper, we add a contribution to this line of research by studying the impact of nucleon superfluid dynamics on the I-Love-Q~relations. 

Apart from newly born neutron stars in supernova explosions and possibly also binary neutron 
stars during the merger phase, typical neutron stars are very cold on the nuclear temperature scale ($\sim10^{10}$ K). 
The internal temperature of a newborn neutron star is expected to drop quickly below the 
transition temperatures ($\sim10^9$ K) for neutrons and protons to become superfluid and superconducting (see, e.g.,~\cite{Lombardo:2001,Haskell:2018,Sedrakian:2019}). 
It is thus expected that nucleon superfluidity will exist in mature neutron stars. In~fact,
the pulsar glitch phenomenon is generally thought to be a manifestation of the neutron superfluid. While the exact mechanism is still not well understood, the~basic idea involves the transfer of angular momentum from the neutron superfluid to the normal component, leading to a sudden spin up. 
In the standard model for large glitches (see~\cite{Haskell:2015} for a review) like those observed 
in the Vela pulsar, the~neutron superfluid rotates by forming a dense array of vortices inside the neutron star. As~the star spins down due to electromagnetic radiation, the~vortices are pinned to the crust~\cite{Anderson:1975} and the neutron superfluid essentially decouples from the normal fluid and does not spin down. 
A lag between the superfluid and the normal component develops, and a Magnus force is induced  
on the vortices. 
When the lag is large enough, the~vortices will suddenly unpin and the superfluid
will spin down. The~crust then spins up, leading to a glitch, due to the conservation of angular momentum. Furthermore, the~long relaxation time following pulsar glitches is also suggested to be a signature of the superfluid component~\cite{Baym:1969}

Since there are strong theoretical and observational motivations to suggest its existence in 
neutron stars, it is important to understand if and how nucleon superfluidity may leave a 
signature on the gravitational waves emitted from a binary neutron star system. After~all, the~stars are still expected to be cold when the system sweeps through the sensitivity bands of 
ground-based detectors. 
It has been estimated that the core temperature of superfluid neutron stars is 
only heated up to $\sim 10^7$ K due to tidal heating in inspiralling binaries~\cite{Yu:2016},
significantly below the superfluid transition temperature ($\sim 10^9$ K). 
However, numerical simulations~\cite{Sekiguchi:2011,Bernuzzi:2016,Perego:2019} suggest that the temperature of a post-merger hypermassive neutron star can be as high as several tens of 
$10^{10}$ K, at which superfluidity is expected to be destroyed. 
As discussed above, the~I-Love-Q relations can be used to reduce the number of matter
parameters in theoretical waveform models. More specifically, it is the Q-Love relation
that is used to express the spin-induced quadrupole moment by the tidal deformability in 
waveform models.  
However, this is possible only if the relations 
are insensitive to different EOS models and physics input. In~this work, we test whether the 
I-Love-Q relations remain valid for superfluid neutron stars. If~the relations are broken by the superfluid dynamics, we may then (in principle) be able to probe the existence and properties of nucleon superfluidity in neutron stars by comparing observational data against waveform models with
and without the assumption of the I-Love-Q~relations.

Protons, electrons, and~nuclei in neutron stars couple strongly (via the electromagnetic 
interaction) on a very short timescale. When neutrons become superfluid, they decouple from the charged components to a first approximation. The~interior of a superfluid neutron star can thus be approximated by a two-fluid system: the neutron superfluid and the ``proton'' fluid containing all charged particles. Besides~being coupled via gravity, the~two fluids can also be coupled
through the entrainment effect so that the flow of one fluid induces a momentum in the other 
(as a result of the strong interaction between neutrons and protons). 
The properties and dynamics of superfluid
neutron stars have been studied using this two-fluid model in a general relativistic framework 
(see~\cite{Andersson:2020} for a review). In~this paper, we use the two-fluid model to study the effects of superfluid dynamics on the I-Love-Q relations. However, we neglect the effects of the
solid crust as it is expected to have only a tiny effect on global stellar quantities like the tidal deformability~\cite{Gittins:2020}. 

The plan of this paper is as follows. In~Section~\ref{sec:2fluid}, we provide an outline of the 
general relativistic two-fluid formalism that we employ to calculate the moment of inertia, 
the spin-induced quadrupole moment, and~the tidal deformability of superfluid neutron stars. 
Section~\ref{sec:eos} describes the EOS models used in this study. Section~\ref{sec:results} presents
our numerical results. Finally, we summarize and discuss our results in Section~\ref{sec:discuss}. 
We use units where $G=c=\hbar=1$ unless otherwise~noted.

%%%%%%%%%%%%%%%%%%%%%%%%%%%%%%%%%%%%%%%%%%
\section{General Relativistic Two-Fluid~Formalism}
\label{sec:2fluid}

%In this section, we shall outline the formulation of our calculation. 
%Readers who are more interested in the final physical results may skip this section and go to 
%Sec.~\ref{sec:results} directly. 

Our study is based on the general relativistic two-fluid formalism developed by
Carter and his collaborators (e.g., \cite{Carter:1989,Comer:1993,Comer:1994,Carter:1998,Langlois:1998}). 
The formalism is built around the master function
$\Lambda(n^2 , p^2 , x^2)$, which is formed by the scalars constructed from the neutron $n^\mu$
and proton $p^\mu$ number density currents: $n^2 = - n_\mu n^\mu$, $p^2 = - p_\mu p^\mu$, and~$x^2 = - n_\mu p^\mu$. 
As already mentioned, the~``proton'' fluid is used to refer to a conglomerate made of all charged 
components. The~master function plays the role of the EOS in the two-fluid formulation. 
Given a master function $\Lambda$, the~stress-energy tensor is determined by
\begin{equation}
T^\mu_\nu = \Psi \delta^\mu_\nu + n^\mu \mu_\nu + p^\mu \chi_\nu  , 
\end{equation}
where the generalized pressure $\Psi$ is given by
\begin{equation}
\Psi = \Lambda - n^\alpha \mu_\alpha - p^\alpha \chi_\alpha  . 
\end{equation}

The chemical potential covectors $\mu_\alpha$ and $\chi_\alpha$ (the fluid four-momenta), respectively, for~the neutrons and (conglomerate) protons are
\begin{equation}
\mu_\alpha = {\cal B} n_\alpha + {\cal A} p_\alpha , \ \ \ 
\chi_\alpha = {\cal C} p_\alpha + {\cal A} n_\alpha ,
\end{equation}
where
\begin{equation}
{\cal A} = - {\partial \Lambda \over \partial x^2} , \ \ \ 
{\cal B} = - 2 {\partial \Lambda \over \partial n^2} , \ \ \
{\cal C} = - 2 {\partial \Lambda \over \partial p^2 } . 
\end{equation}

The coefficient $\cal A$ captures the entrainment effect~\cite{Andersson:2020,Andersson:2021}  between the two fluids through which the current of one fluid will induce a momentum in the other fluid. The~equations of motion for the two fluids consist of two conservation equations,
\begin{equation}
\nabla_\mu n^\mu = 0, \ \ \ 
\nabla_\mu p^\mu = 0, 
\end{equation}
and two Euler equations,
\begin{equation}
n^{\mu} \nabla_{[\mu }\mu_{\nu ] } = 0 , \ \ \
p^{\mu} \nabla_{[\mu } \chi_{\nu ]} = 0 .
\end{equation}

The I-Love-Q relations connect the moment of inertia $I$ and quadrupole moment $Q$ of slowly rotating 
stars to the tidal deformability $\lambda_{\rm tid}$ of nonrotating stars. These quantities are determined by perturbative calculations starting from a nonrotating equilibrium background solution. 
In the following, we shall outline the main steps involved in deriving these quantities and summarize the relevant equations for our discussion. We refer the reader to~\cite{Comer:1999,Andersson:2001,Char:2018} for more~details. 

\subsection{Nonrotating~Stars}

The unperturbed background solution is assumed to be a spherically symmetric and static spacetime described by the metric
\begin{equation}
ds^2 = - e^{\nu(r)} dt^2 + e^{\lambda(r)} dr^2 + r^2 (d\theta^2 + \sin^2 \theta d\phi^2 ) .
\end{equation} 

The equilibrium structure of a nonrotating superfluid neutron star, assuming that the neutron and proton fluids coexist throughout the whole star, is studied in~\cite{Comer:1999}. We also make this 
assumption in our study. 
This is obviously a simplified model as a realistic neutron star is expected to be 
a multilayer system due to the physics of the superfluid phase transition~\cite{Andersson:2002,Lin:2008}. 
For instance, due to the density dependence of the superfluid gap function, the~two-fluid region may be sandwiched by single-fluid layers in a realistic neutron star (see Figure~2 of~\cite{Lin:2008} for an illustration).  
However, we expect that the multilayer aspects will reduce the 
two-fluid region in the star and make the I-Love-Q relations less sensitive to the two-fluid dynamics. As~we shall see later, the~I-Love-Q relations remain valid to high accuracy for our simplified two-fluid stellar model unless the value of the spin ratio between the two fluids becomes quite unrealistic. We expect that a more realistic multilayer model would strengthen this~conclusion. 

The resulting metric and hydrostatic equilibrium equations are given by 
Equations~(25)--(27) of~\cite{Comer:1999}:
\begin{eqnarray}
\lambda^{\prime} = {1 - e^{\lambda} \over r} - 8 \pi r 
                         e^{\lambda} \Lambda , && \cr
\nu^{\prime} = - {1 - e^{\lambda} \over r} + 8 \pi r 
                         e^{\lambda} \Psi , && \cr
 {\cal A}^0_0 p^{\prime} + {\cal B}^0_0 n^{\prime} + {1 \over 2} ( {\cal B} n + {\cal A} p) 
           \nu^{\prime} = 0 , && \cr                        
{\cal C}^0_0 p^{\prime} + {\cal A}^0_0 n^{\prime} + {1 \over 2} ( {\cal A} n + {\cal C} p) 
           \nu^{\prime} = 0 , && 
 \label{eq:backgd}
\end{eqnarray}
where primes denote derivatives with respect to $r$ and the coefficients ${\cal A}^0_0$, ${\cal B}^0_0$, and~${\cal C}^0_0$ are determined from the master function $\Lambda$ 
(see~\cite{Comer:1999} for the explicit expressions). 
The radius $R$ of the background star is defined by the condition that the pressure vanishes at the surface (i.e., $\Psi(R)=0$) and the total mass $M$ is obtained by
\begin{equation}
M = - 4\pi \int^{R}_0 r^2 \Lambda(r) dr . 
\end{equation}

\subsection{Slowly Rotating~Stars}
\label{sec:rotate}

Rotating superfluid neutron stars have been studied using the general relativistic two-fluid 
formalism~\cite{Andersson:2001,Prix:2005,Sourie:2016}. In~this work, we determine the moment
of inertia and quadrupole moment of slowly rotating superfluid neutron stars by following the
formalism developed in~\cite{Andersson:2001}. The~spacetime for a rotating star is given by an axisymmetric and stationary metric of the form
\begin{equation}
ds^2 = - \left[ N^2 - \sin^2 \theta K (N^\phi)^2 \right] dt^2 + V dr^2 - 2 \sin^2 \theta K N^\phi
dt d\phi + K \left( d\theta^2 + \sin^2 \theta d\phi^2 \right) .  
\end{equation}

The neutron and proton fluids are assumed to be uniformly rotating with angular velocities 
$\Omega_{\rm n}$ and $\Omega_{\rm p}$, respectively. In~the slow rotation approximation, the~frequencies
$\left( \Omega_{\rm n} , \Omega_{\rm p} , \sqrt{\Omega_{\rm n} \Omega_{\rm p} } \right)$ are assumed to be much smaller than the characteristic Kepler frequency, which is proportional to $\sqrt{M/R^3}$ in Newtonian theory, and~one can then expand the metric functions up to second order in the rotation as
\begin{eqnarray}
&& N = e^{\nu(r)/2} ( 1 + h_0(r) + h_2(r) P_2 (\cos \theta) ) ,  \cr
&& V = e^{\lambda(r)} ( 1 + 2 v_0 (r) + 2 v_2 (r) P_2 (\cos \theta) ) , \cr
&& K = r^2 ( 1 + 2 k_2 (r) P_2 (\cos \theta) ) , \cr
&& N^\phi = \omega(r) , 
\label{eq:metric_perturb}
\end{eqnarray}
where $P_2 (\cos \theta) = (3 \cos^2 \theta - 1)/2$ is the second-order Legendre polynomial. 
The perturbed metric functions $h_0$, $h_2$, $v_0$, $v_2$, and~$k_2$ are second order in 
angular velocities. The~frame-dragging term $\omega(r)$ is a first-order quantity determined
by
\begin{equation}
{1\over r^4} \left( r^4 e^{-(\lambda + \nu)/2} {\tilde L}^{\prime}_{\rm n} \right)^{\prime}
-16\pi e^{ (\lambda - \nu)/2} (\Psi - \Lambda) {\tilde L}_{\rm n} 
= 16\pi e^{(\lambda - \nu)/2} \chi p (\Omega_{\rm n} - \Omega_{\rm p} ) , 
\label{eq:frame_drag}
\end{equation} 
where ${\tilde L}_{\rm n} = \omega - \Omega_{\rm n}$ and 
${\tilde L}_{\rm p} = \omega - \Omega_{\rm p}$. It should be noted
that the equation reduces to the corresponding frame-dragging equation for a single-fluid
model~\cite{Hartle:1967} when the two fluids are co-rotating (i.e., $\Omega_{\rm n} = \Omega_{\rm p}$). Equation~(\ref{eq:frame_drag}) is solved by choosing a central value 
${\tilde L}_{\rm n} (0)$ so that the interior and exterior solutions match at the stellar surface 
to satisfy the condition
\begin{equation}
{\tilde L}_{\rm n} (R) = - \Omega_{\rm n} + {2 J \over R^3} , 
\end{equation}
where $J = J_{\rm n} + J_{\rm p}$ is the total angular momentum of the star. The~neutron and proton angular momenta are given, respectively, by~\begin{eqnarray} 
&& J_{\rm n} = - {8\pi\over 3} \int^R_0 dr r^4 e^{(\lambda - \nu)/2} \left[ \mu n {\tilde L}_{\rm n} 
+ {\cal A} n p (\Omega_{\rm n} - \Omega_{\rm p} ) \right] , \cr
&& J_{\rm p} = - {8\pi\over 3} \int^R_0 dr r^4 e^{(\lambda - \nu)/2} \left[ \chi p {\tilde L}_{\rm p} 
+ {\cal A} n p (\Omega_{\rm p} - \Omega_{\rm n} ) \right]  .
\end{eqnarray}

The moment of inertia of a single-fluid star in general relativity is defined by the ratio between the angular momentum and angular velocity of the star. We generalize the definition to a two-fluid star so that the moments of inertia of the neutron and proton fluids are defined, respectively, by~\begin{equation}
I_{\rm n} = {J_{\rm n} \over \Omega_{\rm n} } , \ \ \ 
I_{\rm p} = {J_{\rm p} \over \Omega_{\rm p} } . 
\end{equation}
 
The total moment of inertia is then given by $I = I_{\rm n} + I_{\rm p}$, which reduces to the definition for a single-fluid star if the two fluids are co-rotating (see also~\cite{Sourie:2016}). 
This completes the discussion of a slowly rotating two-fluid star up to first order in the angular~velocities. 

In order to study the spin-induced quadrupole moment $Q$ of a slowly rotating star, one needs
to consider the metric and fluid perturbations up to second order in the angular velocities. 
The second-order metric perturbation functions defined in Equation~(\ref{eq:metric_perturb}) consist of the $l=0$ ($h_0 , v_0$) and $l=2$ ($h_2 , v_2 , k_2$) contributions, where $l$ is the order of the Legendre polynomials $P_l (\cos\theta)$. The~equations and numerical schemes for determining the interior solutions of these quantities can be found in~\cite{Andersson:2001}. Outside the star, the~problem is identical to the single-fluid case and the metric functions can be obtained analytically~\cite{Hartle:1967}. In~particular, the~exterior solution for $h_2$ is given by
\begingroup\makeatletter\def\f@size{9}\check@mathfonts
\def\maketag@@@#1{\hbox{\m@th\normalsize\normalfont#1}}%
\begin{eqnarray} 
h_2 (r) &=& - A \left[ {3\over 2} \left( {r\over M}\right)^2 \left(1 - {2M\over r}\right) 
\ln \left( 1 - {2M\over r} \right) 
+ { (r-M)(3 - 6M/r - 2(M/r)^2 ) \over M (1-2M/r) }  \right]    \cr
&&+ {J^2 \over M r^3} \left(1 + {M\over r} \right) ,
\end{eqnarray} 
\endgroup
where the constant $A$ is determined by matching the interior and exterior solutions at the stellar 
surface. Far from the star, the~function $h_2 (r) P_2(\cos\theta)$ becomes the perturbed Newtonian potential and the quadrupole moment can be read off from the coefficient of the 
$ P_2(\cos\theta)/r^3$ term:
\begin{equation}
Q = -{8\over 5} A M^3 - {J^2\over M} .
\end{equation}
    
The normalized (dimensionless) moment of inertia $\bar I$ and quadrupole moment $\bar Q$ that 
appear in the I-Love-Q relations are defined by
\begin{equation}
{\bar I} = {I \over M^3} , \ \ \ 
{\bar Q} = - {Q\over a^2 M^3 } , 
\end{equation}
where $a= J/M^2$ is the dimensionless spin~parameter.

\subsection{Tidally Deformed Nonrotating~Stars}
\label{sec:Love}

\textls[-5]{The tidal deformability $\lambda_{\rm tid}$ measures the deformation of a neutron star due to the 
tidal field produced by the companion in a binary system. 
In the static-tide limit and when the separation between the two stars is large compared to the radii of the stars, the~computation of $\lambda_{\rm tid}$ for nonrotating single-fluid neutron stars 
is well} established~\cite{Hinderer:2008,Damour:2009}. 
The tidal deformations of slowly rotating single-fluid neutron stars have also been studied in~\cite{Pani:2015,Landry:2017}.
The formulation has recently been extended to nonrotating superfluid neutron stars within the general relativistic two-fluid formalism~\cite{Char:2018,Datta:2020}. 
Similar to the study of slowly rotating stars discussed in \mbox{Section~\ref{sec:rotate},} the~tidal deformability is determined by perturbing a nonrotating background solution. We focus on the 
dominant quadrupolar ($l=2$) static tidal field and consider the even-parity perturbations in the Regge--Wheeler gauge so that the full metric is given by
\begin{eqnarray}
ds^2 &=& - e^{\nu(r)} \left[ 1 + H_0 (r) P_2 (\cos\theta) \right] dt^2 + e^{\lambda (r)} 
\left[ 1 + H_2 (r) P_2 (\cos\theta) \right] dr^2    \cr
&& + r^2 \left[ 1 + K(r) P_2 (\cos \theta) \right] ( d\theta^2 + \sin^2\theta d\phi^2 ) .
\end{eqnarray}
 
The perturbed Einstein equations impose the condition $H_0 = - H_2 \equiv H$. The~linearized metric and fluid equations inside the star yield the following equation for determining the tidal deformability~\cite{Char:2018}
\begingroup\makeatletter\def\f@size{9}\check@mathfonts
\def\maketag@@@#1{\hbox{\m@th\normalsize\normalfont#1}}%
\begin{equation} 
H^{''} + H^{'} \left\lbrace {2 \over r} + e^\lambda \left[ {2 m \over r^2} 
+ 4\pi r (\Psi+\Lambda) \right] \right\rbrace 
+ H \left[ -{6e^\lambda \over r^2} + 4\pi e^\lambda (9\Psi - 5\Lambda - g) - \nu^{'2 } 
\right] = 0 ,
\label{eq:Love_H}  
\end{equation}
\endgroup
where the function $m$ is defined by $e^{\lambda(r)} = (1 - 2 m(r)/r )^{-1}$ and
\begin{equation}
g = { { \mu^2 {\cal C}^0_0 + \chi^2 {\cal B}^0_0 - 2\mu \chi {\cal A}^0_0 } \over 
({\cal A}^0_0)^2 - {\cal B}^0_0 {\cal C}^0_0 } .
\end{equation} 

Noticing that $\Psi$ is the pressure and $\Lambda$ is the negative energy density of the unperturbed
nonrotating background star, one can compare Equation~(\ref{eq:Love_H}) with the corresponding equation in a single-fluid situation (see Equation~(15) of~\cite{Hinderer:2008}) for a 
barotropic EOS $P(\rho)$ and observe that the term $(\rho+P)/(dP/d\rho)$ in the single-fluid 
equation is now replaced by the function $-g$ in the two-fluid equation. 
It can also be shown that Equation~(\ref{eq:Love_H}) reduces to Equation~(15) of~\cite{Hinderer:2008} when the master function depends only on one particle density (e.g., $\Lambda(p^2)$) in the single-fluid limit. 
Once the interior problem (Equation~(\ref{eq:Love_H})) is solved for the metric perturbation function 
$H$, the~remaining step to determine the tidal deformability is the same as that for the single-fluid
counterpart. 

\textls[-15]{The response of a nonrotating star to an external quadrupolar tidal field ${\cal E}_{ij}$ results in the following expansion of the metric function $g_{tt}$ in the star's local asymptotic rest frame
~\cite{Hinderer:2008}:}
\begin{equation} 
- {1+g_{tt} \over 2} = -{M \over r} - {3 Q_{ij} \over 2 r^3} \left( {x^i x^j\over r^2}
-{1 \over 3} \delta^{ij} \right) + {1\over 2} {\cal E}_{ij} x^i x^j ,
\label{eq:gtt_expand}
\end{equation}
where $Q_{ij}$ is the tidally induced traceless quadrupole moment tensor of the star. The~tidal deformability $\lambda_{\rm tid}$ of the star is defined by $Q_{ij} = -\lambda_{\rm tid} {\cal E}_{ij}$. It is also convenient to define the dimensionless ($l=2$) tidal Love number $k_2 \equiv 3 \lambda_{\rm tid} /2 R^5$. 
Outside the star, the~solutions of Equation~(\ref{eq:Love_H}) are given by the associated Legendre
functions $Q_2^2(x)$ and $P_2^2(x)$:
\begin{equation}
H(r) = c_1 Q_l^2(x) + c_2 P_l^2(x) ,
\end{equation}
where $x = r/M - 1$. Using Equation~(\ref{eq:gtt_expand}), the~coefficients $c_1$ and $c_2$ are related
to the tidal Love number by $k_2 = 4c_1 M^5 / 15 c_2 R^5$. The~value of $c_1/c_2$ can be obtained by matching the interior and exterior solutions of Equation~(\ref{eq:Love_H}) at the surface. The~tidal Love number can then be expressed as~\cite{Hinderer:2008}
\begingroup\makeatletter\def\f@size{9}\check@mathfonts
\def\maketag@@@#1{\hbox{\m@th\normalsize\normalfont#1}}%
\begin{eqnarray}
  k_2 &=&  {8\over 5} C^5 (1-2C)^2 \left[2 + 2C\left(y - 1\right) - y\right]
   \Bigg\{2C\left[4(1+y)C^4 + (6y-4)C^3 + (26-22y)C^2  \right. \nonumber \\
   && \left. +3C(5y-8) -3y+6 \right]
    + 3(1-2C)^2\left[2 - y + 2C (y - 1)\right] \text{log}(1 - 2C)\Bigg\}^{-1},
\end{eqnarray}  
\endgroup             
where $C = M/R$ is the compactness and $y = R H^{'}(R)/H(R)$. Finally, the~normalized (dimensionless) tidal deformability ${\bar \lambda}_{\rm tid}$ that appears in the I-Love-Q relations is defined by
\begin{equation} 
{\bar \lambda}_{\rm tid} = { \lambda_{\rm tid} \over M^5} =  {2 \over 3} k_2 C^{-5} . 
\end{equation}

\section{Equations Of~State}
\label{sec:eos}

\subsection{Two-Fluid Polytropic~Model}

As discussed in the previous section, the~master function $\Lambda(n^2 , p^2, x^2)$ plays the role 
of EOS in the two-fluid formulation. We will use two different EOS models in this work. The~master
function of the first model is taken to be~\cite{Comer:1999}
\begin{equation} 
\Lambda = - m_{\rm n} n - \sigma_{\rm n} n^{\beta_{\rm n} } - m_{\rm p} p - 
\sigma_{\rm p} p^{\beta_{\rm p} } , 
\end{equation}  
where the neutron and proton masses are assumed to be equal ($m_{\rm n} = m_{\rm p}$) for simplicity. 
We choose the parameters $\sigma_{\rm n} = 0.2 m_{\rm n}$, $\sigma_{\rm p} = 2 m_{\rm n}$, 
$\beta_{\rm n} = 2.3$, and~$\beta_{\rm p} = 1.95$ as in~\cite{Andersson:2001} to construct background equilibrium models in this work. 
As the master function does not depend on $x^2$, there is no entrainment effect and the fluids 
behave as independent polytropes and couple only through gravity. The~various coefficients ${\cal A}, {\cal B}$, etc., (see Equation~(\ref{eq:backgd})) can be easily computed for this model~\cite{Comer:1999}. In~particular, we have ${\cal A}={\cal A}^0_0 = 0$. 

\subsection{Relativistic Mean Field~Model} 
\label{sec:RMF}

The second EOS that we will consider is a more realistic model based on relativistic mean field (RMF) approximation. RMF based models have been used previously in the study of general relativistic superfluid neutron stars.
Comer and Joynt~\cite{Comer:2003,Comer:2004} have calculated the entrainment effect in a relativistic $\sigma-\omega$ model. Kheto and Bandyopadhyay~\cite{Kheto:2014} later extended the study to include the isospin dependence by using a relativistic $\sigma-\omega-\rho$ model. 
More recently, Char and Datta~\cite{Char:2018,Datta:2020} used the same RMF model to study the tidal deformability of superfluid neutron stars. In~this work, we also use the relativistic $\sigma-\omega-\rho$ model to construct a master function. The~Lagrangian density
for the system is given by
\begin{eqnarray}
{\cal L} &=& \sum_{B={\rm n,p} } {\bar \Psi}_{B} \left (i \gamma_\mu \partial^{\mu} - m_{B} + g_{\sigma {B} } \sigma - g_{\omega {B}} \gamma_\mu \omega^{\mu} - g_{\rho {B}} 
\gamma_\mu \boldsymbol{\tau}_{B} \cdot 
\boldsymbol{\rho}^\mu  \right) {\Psi}_{B}   \cr
&& - {1\over 2} \partial_\mu \sigma \partial^\mu \sigma -{1\over 2} m^2_\sigma \sigma^2
-{1\over 3} b m (g_\sigma \sigma)^3 - {1\over 4} c (g_\sigma \sigma)^4 
-{1\over 4} \omega_{\mu \nu} \omega^{\mu \nu} \cr
&& - {1\over 2} m^2_{\omega} \omega_\mu \omega^\mu - 
{1\over 4} \boldsymbol{\rho}_{\mu\nu} \cdot \boldsymbol{\rho}^{\mu\nu}
-{1\over 2} m^2_\rho \boldsymbol{\rho}_\mu \cdot \boldsymbol{\rho}^\mu , 
\end{eqnarray}
where $\Psi_{B}$ is the Dirac spinor for baryons ${B}$ with baryon mass $m_{B}$ and $\gamma_\mu$ denotes the Dirac matrices. The~isospin operator is denoted by $\boldsymbol{\tau}_B$. 
In this theory, we have the self-interacting scalar $\sigma$ field, the~vector omega field $\omega^\mu$, and~the isovector rho field $\boldsymbol{\rho}^\mu$. 
The latter two fields have the corresponding field tensors 
$\omega_{\mu \nu}$ and $\boldsymbol{\rho}_{\mu\nu}$. The~nucleon mass $m$ is taken to be the average of the bare neutron and proton masses. 
The resulting equations of motion from the Lagrangian are solved 
in the RMF approximation. The~entrainment effect is incorporated by choosing a frame in which the neutrons have zero spatial momentum and the proton momentum is given by
$k_{\rm p}^\mu =(k_0 , 0, 0, K)$ \cite{Comer:2003}. In~the limit $K \rightarrow 0$ that is relevant to the slow rotation approximation considered in this work, the~master function for determining the 
equilibrium background neutron stars is given explicitly by~\cite{Char:2018}
\begin{eqnarray}
\Lambda &=& - {c_\omega^2\over 18\pi^4} (k_{\rm n}^3 + k_{\rm p}^3)^2 - { c_\rho^2 \over 72\pi^4}
(k_{\rm p}^3 - k_{\rm n}^3)^2 - {1\over 4\pi^2} \left( k_{\rm n}^3 
\sqrt{ k_{\rm n}^2 + m_*^2 |_0 } +k_{\rm p}^3 \sqrt{ k_{\rm p}^2 + m_*^2 |_0 }  \right) \cr
&&  -{1\over 4 c_\sigma^2 } \Big\{ (2m - m_* |_0 ) (m - m_* |_0 ) 
+ m_* |_0 \left[  b m c_\sigma^2 ( m - m_* |_0 )^2 
+ c c_\sigma^2 ( m - m_* |_0 )^3 \right] \Big\}  \cr
&& - {1\over 3} b m ( m - m_* |_0 )^3 - {1\over 4} c ( m - m_* |_0 )^4 
- {1\over 8\pi^2} \Bigg\{ k_{\rm p} ( 2k_{\rm p}^2 + m_{\rm e}^2 ) \sqrt{ k_{\rm p}^2 + m_{\rm e}^2}  \cr
&&  - m_{\rm e}^4 \ln \left( { k_{\rm p} + \sqrt{k_{\rm p}^2 + m_{\rm e}^2} \over m_{\rm e} } \right) \Bigg\} ,
\end{eqnarray}
where $k_{\rm n} = (3\pi^2 n)^{1/3}$ and $k_{\rm p} = (3\pi^2 p)^{1/3}$ are the neutron and proton Fermi momenta, respectively. We have also added the contributions due to the electrons in the master
function (the last term containing the electron mass $m_{\rm e}$). 
The background value for the Dirac effective mass $m_* |_0$ is determined by the following transcendental equation
\begin{eqnarray}
m_* |_0 &=& m - m_* |_0 {c_\sigma^2 \over 2\pi^2} \Bigg\{ k_{\rm n} \sqrt{k_{\rm n}^2 + m^2_* |_0 }
+ k_{\rm p} \sqrt{k_{\rm p}^2 + m^2_* |_0 }  \cr
&& + {1\over 2} m^2_* |_0 \ln \left[ 
{ -k_{\rm n} + \sqrt{k_{\rm n}^2 + m^2_* |_0 } \over k_{\rm n} + \sqrt{k_{\rm n}^2 + m^2_* |_0 } } \right]  +{1\over 2} m^2_* |_0 \ln \left[ 
{ -k_{\rm p} + \sqrt{k_{\rm p}^2 + m^2_* |_0 } \over k_{\rm p} + \sqrt{k_{\rm p}^2 + m^2_* |_0 } } \right] \Bigg\}  \cr
&& + b m c_\sigma^2 (m - m_* |_0 )^2 + c c_\sigma^2 (m - m_* |_0 )^3 . 
\end{eqnarray} 

%\begingroup\makeatletter\def\f@size{8.5}\check@mathfonts
%\def\maketag@@@#1{\hbox{\m@th\normalsize\normalfont#1}}%
%\begin{eqnarray}
%m_* |_0 &=& m - m_* |_0 {c_\sigma^2 \over 2\pi^2} \Bigg\{ k_{\rm n} \sqrt{k_{\rm n}^2 + m^2_* |_0 }
%+ k_{\rm p} \sqrt{k_{\rm p}^2 + m^2_* |_0 } + {1\over 2} m^2_* |_0 \ln \left[ 
%{ -k_{\rm n} + \sqrt{k_{\rm n}^2 + m^2_* |_0 } \over k_{\rm n} + \sqrt{k_{\rm n}^2 + m^2_* |_0 } } %\right]  \cr
%&&  +{1\over 2} m^2_* |_0 \ln \left[ 
%{ -k_{\rm p} + \sqrt{k_{\rm p}^2 + m^2_* |_0 } \over k_{\rm p} + \sqrt{k_{\rm p}^2 + m^2_* |_0 } } %\right] \Bigg\} + b m c_\sigma^2 (m - m_* |_0 )^2 + c c_\sigma^2 (m - m_* |_0 )^3 . 
%\end{eqnarray} 
%\endgroup

We refer the reader to~\cite{Char:2018} for the expressions of the various coefficients $\cal A$, 
$\cal B$, etc. (see Equation~(\ref{eq:backgd})). In~contrast to the two-fluid polytropic model, it 
should be noted that the coefficient $\cal A$, which is responsible for the entrainment effect, 
is nonzero in this RMF model. 
The coupling constants $c_\sigma^2 \equiv (g_\sigma / m_\sigma)^2$, $c_\omega^2 \equiv (g_\omega 
/ m_\omega)^2$, $c_\rho^2 \equiv (g_\rho / m_\rho)^2$, $b$, and~$c$ in the master function are determined by the nuclear matter saturation properties~\cite{Kheto:2014}. 
In the following, we shall use the same NL3~\cite{Fattoyev:2010}
and GM1~\cite{Glendenning:1991} parameter sets (see Table~\ref{tab:eos}) as employed in~\cite{Char:2018}.

\begin{specialtable}
\caption{Parameter sets for the NL3 and GM1 models. The~nucleon-meson coupling constants
$c_\sigma^2$, $c_\omega^2$, and~$c_\rho^2$ are expressed in ${\rm fm}^2$, while $b$ and $c$
are dimensionless~\cite{Char:2018}. }
%\centering
\begin{tabular*}{\hsize}{@{}@{\extracolsep{\fill}}cccccc@{}}
\toprule
\textbf{Model}	& \boldmath{$c_\sigma^2$}	& \boldmath{$c_\omega^2$} & \boldmath{$c_\rho^2$} & \boldmath{$b$} & \boldmath{$c$} \\
\midrule
NL3		& 15.739  &  10.530  &  5.324  &  0.002055 	& $-0.002650$  \\
GM1		& 11.785  &  7.148   &  4.410  &  0.002948  &  $-0.001071$   \\
\bottomrule
\end{tabular*}
\label{tab:eos}
\end{specialtable}

\begin{figure}[t]
  {\includegraphics[width=300pt]{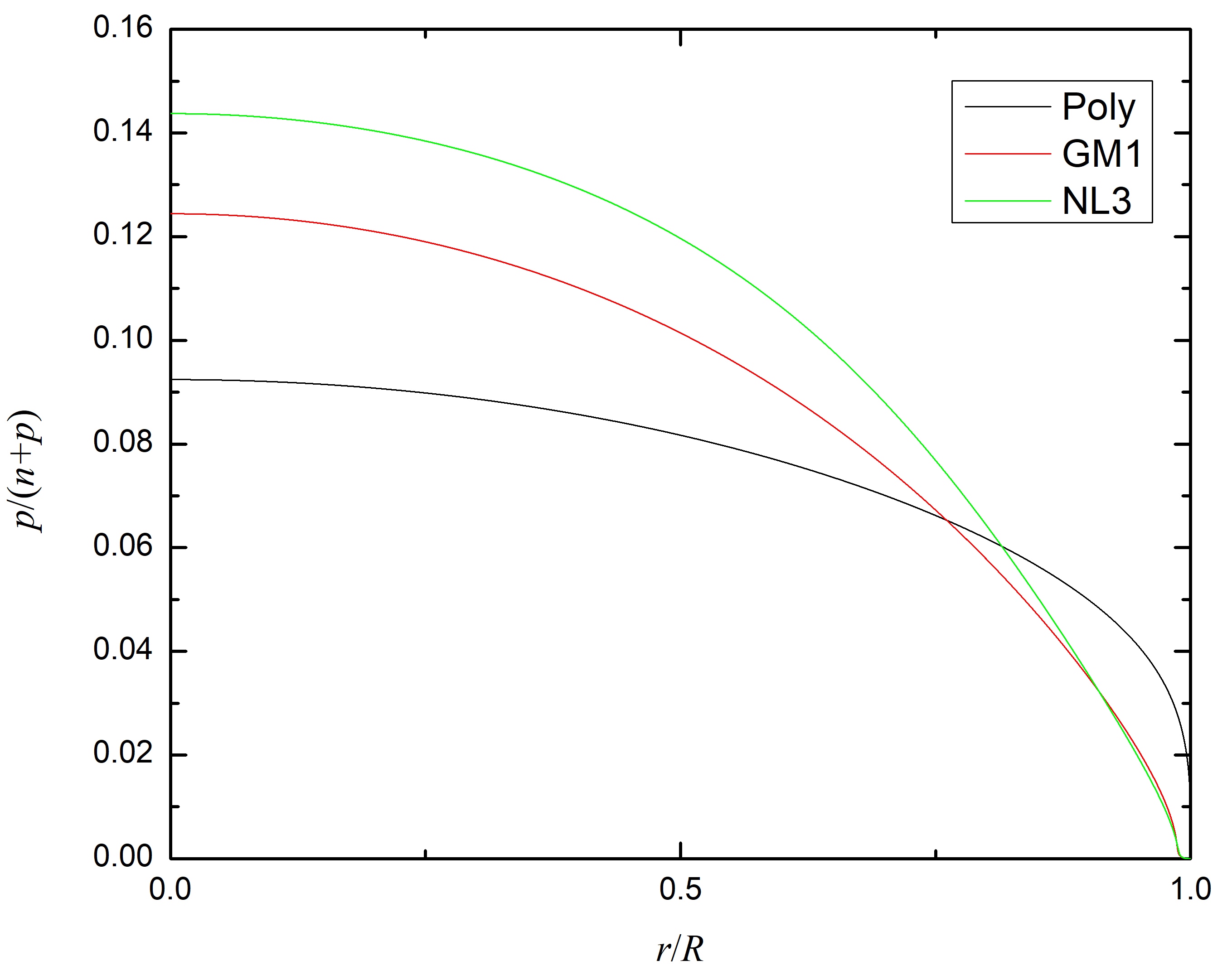}}
  \caption{Profiles of the proton fraction $p/(n+p)$ for $1.4 M_\odot$ neutron stars constructed 
  from the polytropic (Poly), GM1, and~NL3 EOS models.     }
  \label{fig:Proton_fraction}
\end{figure}

Before studying the I-Love-Q relations, it is worth pointing out that our two-fluid stellar models are dominated by the neutron superfluid as the typical central value of the proton fraction in our
models is about 10\%, which is comparable to that of a realistic neutron star. For~example, Figure~\ref{fig:Proton_fraction} shows the profiles of the proton fraction $p/(n+p)$ for $1.4 M_\odot$ neutron stars constructed from our EOS~models.

\unskip

\section{Numerical~Results}
\label{sec:results}

The I-Love-Q universal relations~\cite{Yagi:2013a,Yagi:2013b} originally found for single-fluid neutron stars connect the three dimensionless quantities $\bar I$, $\bar Q$, and~
${\bar \lambda}_{\rm tid}$. 
While one might naively expect that $\bar I$ and $\bar Q$ are somehow related to each other as 
they both characterize the effects of rotation, the~fact that these quantities together with 
${\bar \lambda}_{\rm tid}$ are connected by the following approximately EOS-independent relations 
are quite surprising~\cite{Yagi:2013b}:
\begin{equation}
\ln y_i = a_i + b_i \ln x_i + c_i (\ln x_i )^2 + d_i (\ln x_i)^3 + e_i (\ln x_i)^4 ,
\label{eq:I-Love-Q}
\end{equation}
where $(x_i , y_i)$ are any two of $\bar I$, $\bar Q$, and~
${\bar \lambda}_{\rm tid}$. The~set of fitting coefficients $(a_i, b_i, c_i, d_i, e_i)$ are different for different pairs of $x_i$ and $y_i$ (see Table I in~\cite{Yagi:2013b}).  

Our aim in this work is to study whether the I-Love-Q relations remain valid for superfluid neutron stars. We have generalized the definitions of $\bar I$, $\bar Q$, and~${\bar \lambda}_{\rm tid}$ 
to two-fluid stars in Section~\ref{sec:2fluid}. 
For a given master function (i.e., a~EOS model) in the two-fluid formulation, the~procedure
in our calculation is to first build a nonrotating background star for given central number densities $n(0)$ and $p(0)$. Note that the two densities are not independent as the background star is assumed to be in chemical equilibrium~\cite{Comer:1999}. A~slowly rotating model is then built by solving the perturbative equations with the nonrotating background model as input. The~neutron
and proton fluids are assumed to be rigidly rotating with angular velocities $\Omega_{\rm n}$ and 
$\Omega_{\rm p}$. However, instead of providing $\Omega_{\rm n}$ and $\Omega_{\rm p}$ as input parameters, the~system of equations can be scaled in such a way that it is sufficient to specify the spin ratio $\Omega_{\rm n} / \Omega_{\rm p}$ in the calculation~\cite{Andersson:2001}. We will thus present our results for stellar sequences specified by $\Omega_{\rm n} / \Omega_{\rm p}$. 
The spin ratio is expected to be very close to unity as suggested by the pulsar glitch 
phenomenon (see Section~\ref{sec:discuss} for discussion). However, we will explore a much wider parameter range to study the validity of the I-Love-Q relations. 
The dimensionless moment of inertia $\bar I$ and quadrupole moment $\bar Q$ can be determined for the slowly rotating star as outlined in Section~\ref{sec:rotate}. The~computation of the dimensionless tidal deformability ${\bar \lambda}_{\rm tid}$ for the nonrotating star was also discussed in Section~\ref{sec:Love}.

\begin{figure}[t]
\widefigure
\begin{minipage}{.5\linewidth}
%\centering
\subfloat[]{\includegraphics[width=8.26cm]{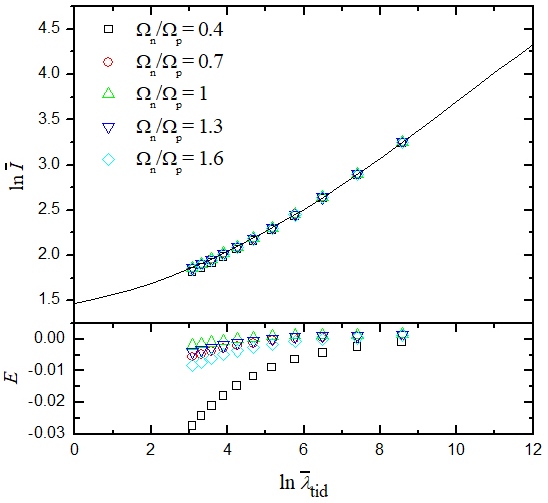}}
\end{minipage}%
\begin{minipage}{.5\linewidth}
%\centering
\subfloat[]{\includegraphics[width=8.26cm]{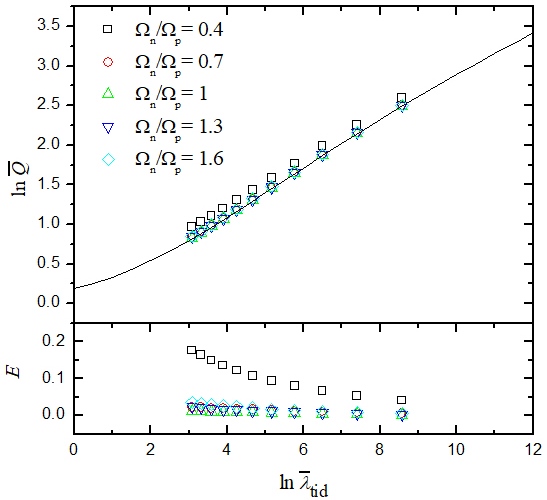}}
\end{minipage}\par\medskip
%\centering
\subfloat[]{\includegraphics[width=8.26cm]{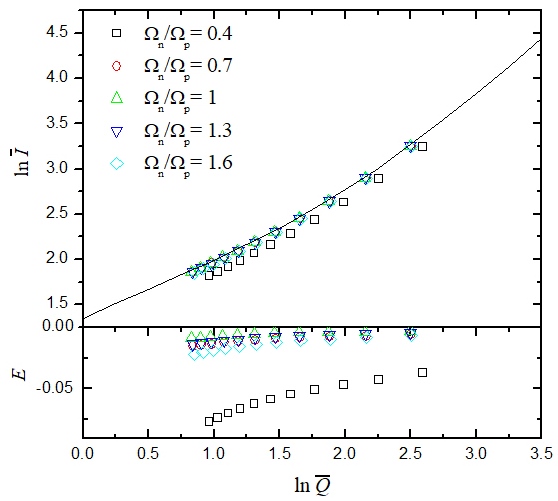}}

\caption{The accuracy of the I-Love (a), Q-Love (b), and I-Q (c) relations for our chosen two-fluid polytropic EOS using different spin ratios $\Omega_{\rm n} / \Omega_{\rm p}$. 
The~solid line in the upper panel of each figure 
represents the corresponding fitting curve for ordinary single-fluid neutron stars~\cite{Yagi:2013b}. 
The lower panel of each figure shows the relative error between the two-fluid results and the 
corresponding fitting curve. }
\label{fig:ILoveQ_poly}
\end{figure}

We first consider the results for the two-fluid polytropic EOS model. In~Figure~\ref{fig:ILoveQ_poly}a, we plot in the upper panel $\ln {\bar I}$ against 
$\ln {\bar \lambda}_{\rm tid}$ (i.e., the~I-Love relation) for sequences of stars with 
spin ratios $\Omega_{\rm n} / \Omega_{\rm p} = 0.4$, 0.7, 1, 1.3, and~1.6.
The solid line is the fitting curve (Equation~(\ref{eq:I-Love-Q})) for the I-Love relation for 
single-fluid ordinary neutron stars~\cite{Yagi:2013b}. The~lower panel shows the relative error, 
$E = ({\hat y} - y)/y$, between~the numerical data, $\hat y$, and~the fitting curve, $y$. 
We see from the upper panel that the numerical data can still match the fitting curve for 
single-fluid stars very well. However, the~effects of two-fluid dynamics become more apparent in the lower panel.
The co-rotating case $\Omega_{\rm n}/\Omega_{\rm p} = 1$ corresponds effectively to a single-fluid system and has the smallest $|E|$ among the different sequences, as~expected. 
It can also be seen that $|E|$ becomes larger as the spin ratio deviates more away from unity. 
For the cases where the neutron fluid rotates faster than the proton fluid 
(i.e., $\Omega_{\rm n}/\Omega_{\rm p} > 1$), the~error $|E|$ increases with 
$\Omega_{\rm n}/\Omega_{\rm p}$. 
On the other hand, $|E|$ increases as $\Omega_{\rm n}/\Omega_{\rm p}$ decreases from 0.7 to 0.4 for the opposite situation where the proton fluid rotates faster.
Since ${\bar \lambda}_{\rm tid}$ decreases with increasing compactness, the~numerical results 
thus illustrate that the error $|E|$ increases with the compactness for a given~sequence.

% start a new page without indent 4.6cm
%\clearpage
\end{paracol}
\nointerlineskip
\begin{paracol}{2}
%\linenumbers
\switchcolumn

In Figure~\ref{fig:ILoveQ_poly}b,c, we plot $\ln {\bar Q}$ against 
$\ln {\bar \lambda}_{\rm tid}$ and $\ln {\bar I}$ against $\ln {\bar Q}$, respectively. 
Similar to Figure~\ref{fig:ILoveQ_poly}a, the~solid lines in these figures are the corresponding
fitting curves for the Q-Love and I-Q relations for single-fluid stars~\cite{Yagi:2013b}. 
The relative errors between the numerical data and the fitting curves are also plotted in the lower panel of the figures. 
Similar to the I-Love relation in Figure~\ref{fig:ILoveQ_poly}a, the~errors $|E|$ for these two 
relations also increase as the spin ratio deviates from unity. 
It should be pointed out that the I-Love-Q relations (i.e., the~solid lines) for single-fluid stars are approximately EOS independent to within about the 1\% level. 
This means that the I-Love-Q relations may be considered to be broken by the effects of two-fluid 
dynamics only for cases where $|E| > 0.01$. 
For our chosen values of $\Omega_{\rm n} / \Omega_{\rm p}$, we see that the I-Love-Q relations are broken only for the case $\Omega_{\rm n}/\Omega_{\rm p} = 0.4$. Taking our finding at face value, it implies that accurate independent measurements of any pair of quantities in the I-Love-Q relations can in principle be used to test the existence of two-fluid dynamics in superfluid neutron stars,
though quite extreme values of $\Omega_{\rm n}/\Omega_{\rm p}$ are needed.
The Q-Love relation plotted in Figure~\ref{fig:ILoveQ_poly}b is the most promising in this 
aspect as the deviations between the two-fluid data and the fitting curve are more significant
than the other two relations. It should be noted that the error for the Q-Love relation can reach up 
to $|E| \sim 0.1$ level for the case $\Omega_{\rm n} / \Omega_{\rm p} = 0.4$.

As discussed in Section~\ref{sec:intro}, we can make use of various universal relations to reduce 
the matter parameters in theoretical waveform models. In~particular, the~Q-Love relation can be used to get rid of $\bar Q$ by expressing it in terms of ${\bar \lambda}_{\rm tid}$, assuming that the
relation is robust and insensitive to EOS and various physics inputs to high accuracy. 
If the effects of two-fluid dynamics can lead to $O(10\%)$ deviation of the Q-Love
relation as illustrated in our results, then this relation should be used with care in waveform 
modeling. This also suggests that not imposing the Q-Love relation in waveform modeling
can, in principle, allow the tests of superfluid dynamics using gravitational wave observations. 
These implications depend clearly on whether the spin ratio $\Omega_{\rm n}/\Omega_{\rm p}$ can deviate significantly from unity, as considered by us in Figure~\ref{fig:ILoveQ_poly}a--c. 
We will return to this issue and assess the astrophysical relevance of our results in 
Section~\ref{sec:discuss}. 

% start a new page without indent 4.6cm
\clearpage
\end{paracol}
\nointerlineskip
\begin{figure}[H]
\widefigure
\begin{minipage}{.5\linewidth}
%\centering
\subfloat[]{\includegraphics[width=8.26cm]{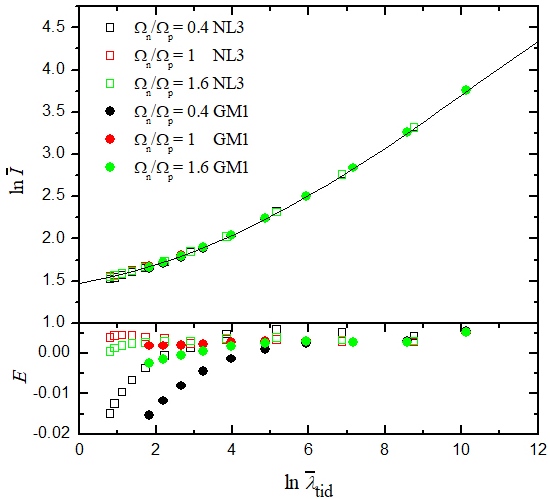}}
\end{minipage}%
\begin{minipage}{.5\linewidth}
%\centering
\subfloat[]{\includegraphics[width=8.26cm]{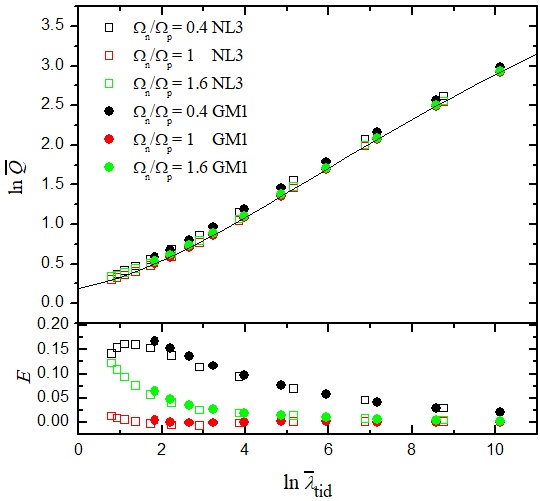}}
\end{minipage}\par\medskip
%\centering
\subfloat[]{\includegraphics[width=8.26cm]{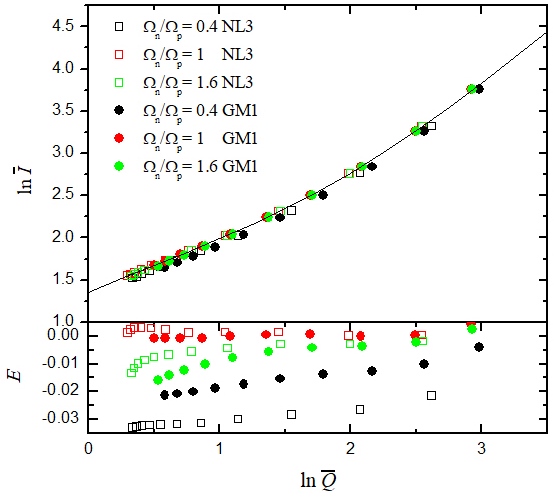}}

\caption{Similar to Figure~\ref{fig:ILoveQ_poly}, but~for the GM1 and NL3 EOS~models. }
\label{fig:ILoveQ_rmf}
\end{figure}
\begin{paracol}{2}
%\linenumbers
\switchcolumn

While we have only presented the results of one particular set of parameters for the two-fluid polytropic EOS model in Figure~\ref{fig:ILoveQ_poly}a--c, we have checked that the results obtained from different sets of EOS parameters are qualitatively the same. 
In particular, the~I-Love-Q relations remain valid to good accuracy unless the spin ratio between the two fluids deviates significantly from unity. However, the~two-fluid polytropic model can 
at best only provide a crude approximation to the properties of superfluid neutron stars since 
the two fluids can couple only through gravity in this model. It is not obvious that the conclusions
derived from this simple model can be generalized directly to more realistic EOS models.  
We fill this gap by considering the RMF model. This model is 
more realistic in the sense that nucleon--nucleon interactions are taken into account through the exchange of effective mesons and the coupling parameters are determined by fitting to known nuclear matter properties. 
The RMF model also has the advantage that the corresponding master function is simple enough that
the various thermodynamics coefficients that we need in the two-fluid calculations can be obtained analytically. Most importantly, in~contrast to the polytropic EOS, the~RMF model contains the 
entrainment effect characterized by ${\cal A} \neq 0$, which is a unique property of superfluid dynamics. 
In Figure~\ref{fig:ILoveQ_rmf}a--c, we present the numerical results for the NL3 and GM1 
parametrizations of the RMF model (see Table~\ref{tab:eos}). 
In each figure, we consider three sequences of $\Omega_{\rm n}/\Omega_{\rm p} = 0.4$, 1, and~1.6 for 
each EOS. We see that the results of NL3 and GM1 EOSs match the I-Love-Q relations to high accuracy for the co-rotating case $\Omega_{\rm n}/\Omega_{\rm p}=1$ as expected. 
The data trends are qualitatively the same as those of the polytropic EOS. 
We note that the relative errors $|E|$ for all three relations increase as the spin ratio deviates away from unity. For~a given sequence of fixed EOS and spin ratio, the~errors also generally 
increase with the compactness. It can also be seen from Figure~\ref{fig:ILoveQ_rmf}b that
the error for the Q-Love relation is the largest among the three relations and can reach up to 
the $|E| \sim 0.1$ level.

\begin{figure}[t]
 {\includegraphics[width=380pt]{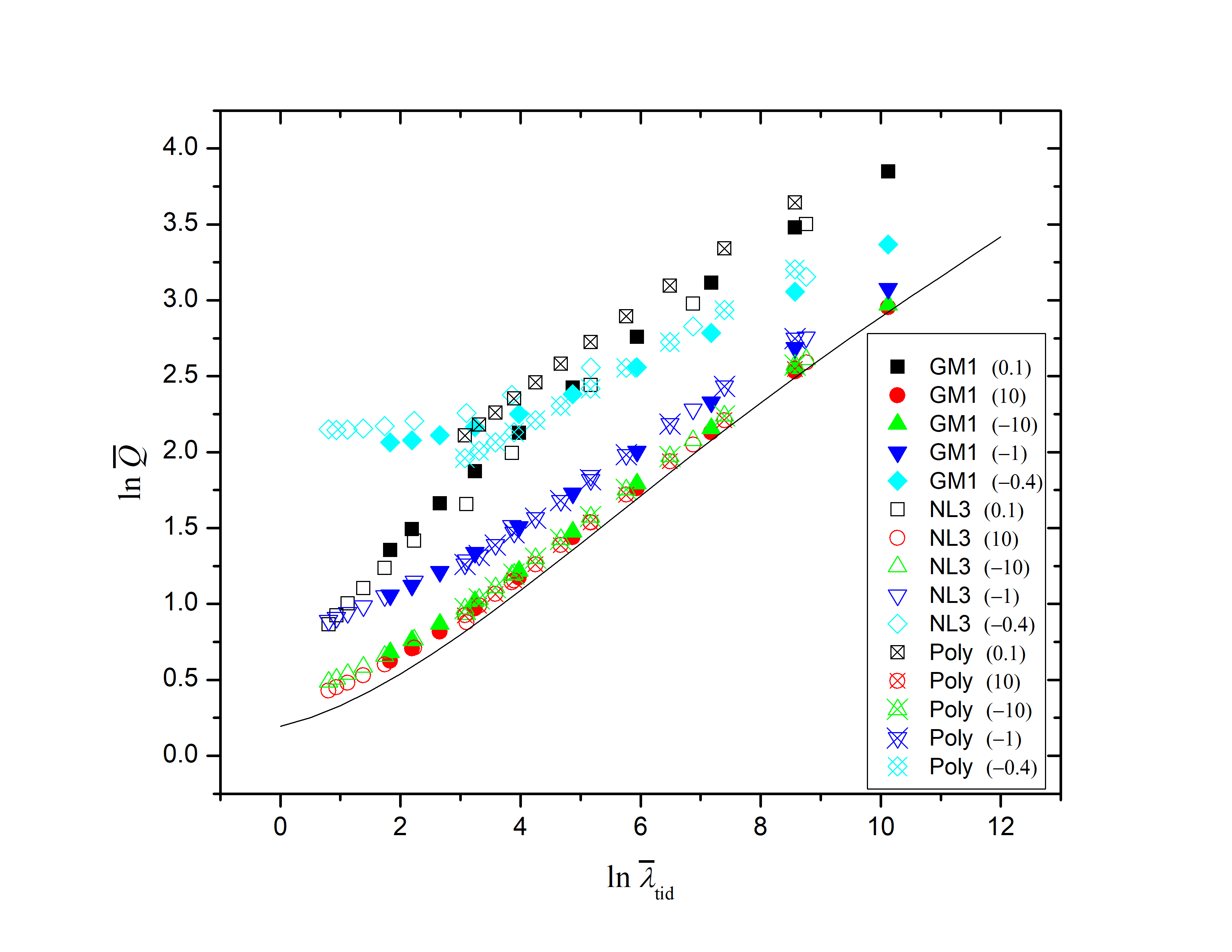}}
  \caption{The breaking of the Q-Love relation for extreme spin ratios as illustrated by the 
  polytropic (Poly), GM1, and~NL3 EOS models. The~numerical values inside the parentheses 
  in the legend box denote the values of $\Omega_{\rm n} / \Omega_{\rm p}$. 
  The solid line is the universal Q-Love relation for ordinary single-fluid neutron stars~\cite{Yagi:2013b}.  }
  \label{fig:QLove_extreme}
\end{figure}  
 
Let us now consider more extreme situations to explore the breaking of 
the Q-Love relation when $\Omega_{\rm n} / \Omega_{\rm p}$ deviates significantly from unity. 
We have studied the trend of the numerical results in the 
$\ln {\bar Q} - \ln {\bar \lambda}_{\rm tid}$ plane. 
In Figure~\ref{fig:QLove_extreme}, we plot the results for a few representative spin ratios 
as an illustration. The~solid line in the figure represents the 
original Q-Love relation for single-fluid stars. We now consider how the data behave as the
spin ratio deviates from unity along the two opposite directions 
$\Omega_{\rm n} / \Omega_{\rm p} > 1$ and $<1$. In~the former case, we find that the results converge 
to a large spin-ratio limit, which can be approximated well by the case 
$\Omega_{\rm n} / \Omega_{\rm p} = 10$ (red data points). Increasing the spin ratio further (e.g., 
$\Omega_{\rm n} / \Omega_{\rm p} = 100$) has little effect on the results. 
In this limit, the~fluid motion is dominated by the neutron superfluid as the proton fraction is 
small (see Figure~\ref{fig:Proton_fraction}).
The possible range of deviation from the Q-Love relation for $\Omega_{\rm n} / \Omega_{\rm p} > 1$ 
is thus approximately bounded by the solid line and the red data points. 
Note, however, that the results for a given spin ratio are still EOS-insensitive in this~range.

We now turn to the situation for $\Omega_{\rm n} / \Omega_{\rm p} < 1$, which turns out to be more interesting. As~we decrease the spin ratio from unity, the~results initially deviate from the
solid line gradually. As~we have seen in Figure~\ref{fig:ILoveQ_rmf}b, the~results for 
$\Omega_{\rm n} / \Omega_{\rm p} = 0.4$ are still quite close to the solid line. However, as~
the spin ratio is decreased further, the~results deviate from the solid line more significantly. We also see that the results become more sensitive to the EOS models as the data shift further away from the solid line. Consider for example the results for $\Omega_{\rm n} /\Omega_{\rm p} = 0.1$ (black data points) in the figure. If~we keep decreasing the spin ratio toward zero, the~data move further upward in the figure
and depend more on the EOS models. However, if~the spin ratio is further decreased to become negative so that the two fluids are counter-rotating, the~data would then shift downward in the figure. The~results for $\Omega_{\rm n} / \Omega_{\rm p} = -0.4$ and $-1$ in the figure show that the results move closer to the solid line and become less sensitive to the EOS models as the spin ratio becomes more negative.  
However, the~results would not converge to the solid line as the spin ratio is decreased further. 
Instead, the~data converge to the large spin-ratio limit that we discussed above for 
$\Omega_{\rm n} / \Omega_{\rm p} > 0$. 
It can be seen from the results of $\Omega_{\rm n} / \Omega_{\rm p} = -10$ (green 
data points) that the data are insensitive to the EOS models and agree quite well with the 
results of $\Omega_{\rm n} / \Omega_{\rm p} =10$. In~fact, this should be expected as the system is 
dominated by the neutron superfluid if $|\Omega_{\rm n} / \Omega_{\rm p} | \gg 1$ and the spin-induced quadrupole moment $Q$ should be independent of the sense of rotation of the neutron superfluid. There should thus be a single large spin-ratio~limit.

Figure~\ref{fig:QLove_extreme} shows that the results become more sensitive to the EOS models as the 
data are far away from the single-fluid universal relation. While it is beyond the scope of 
this work to study the EOS dependence for these extreme cases in general, we have used different parameters for the two-fluid polytropic model to gain some understanding of the general trend. 
In Table~\ref{tab:poly}, we compare the values of $\ln {\bar Q}$ at 
$\ln {\bar \lambda}_{\rm tid} = 5$ for three different polytropic EOS models. 
The EOS parameters are chosen in such a way that the central proton fractions for $1.4 M_\odot$ neutron stars are 0.02 (Poly\_2), 0.04 (Poly\_4), and~0.08 (Poly\_8). For~comparison, our default polytropic model used in Figure~\ref{fig:QLove_extreme} has a central proton fraction about 0.09 for 
a $1.4 M_\odot$ star (see Figure~\ref{fig:Proton_fraction}). In~the table, the~numerical values inside the parentheses are the percentage differences between the data and the value of $\ln {\bar Q}$ predicted by the single-fluid Q-Love relation. 
For a given EOS model, we see that the percentage differences increase significantly as the spin 
ratio decreases from 0.2 to 0.1. This is consistent with the above observation that the deviations
of the superfluid data from the I-Love-Q relations increase as the spin ratio deviates further away
from unity. On~the other hand, the~deviations decrease with the proton fraction inside the stars
for a fixed spin ratio. This agrees with the expectation that the effects of two-fluid dynamics 
should become less important if the system is dominated by one of the~fluids.    

\begin{specialtable}[H]
\caption{Values of $\ln {\bar Q}$ at $\ln {\bar \lambda}_{\rm tid}=5$ for three 
different polytropic EOS models with spin ratios 0.1 and 0.2. The~EOS parameters are chosen in 
such a way that the central proton fractions for $1.4 M_\odot$ stellar models are 0.02 (Poly\_2), 
0.04 (Poly\_4), and~0.08 (Poly\_8). 
The numerical values inside the parentheses are the percentage differences between the data and 
the predictions of the single-fluid Q-Love relation.  }
%\centering
\begin{tabular*}{\hsize}{@{}@{\extracolsep{\fill}}cccc@{}}
\toprule
 \boldmath{$\Omega_{\rm n}/\Omega_{\rm p}$} & \textbf{Poly\_2} & \textbf{Poly\_4} &\textbf{Poly\_8 } \\
\midrule
0.1		& 2.22 (59\%)  &  2.46 (76\%)  &  2.66 (91\%)   \\
0.2	    & 1.66 (18\%)  &  1.79 (28\%)  &  1.96 (40\%)     \\
\bottomrule
\end{tabular*}
\label{tab:poly}
\end{specialtable}

\textls[-5]{The results presented in Figure~\ref{fig:QLove_extreme} may have little astrophysical relevance
as $|\Omega_{\rm n} / \Omega_{\rm p}|$ is not expected to deviate from unity significantly. It is also uncertain how the two fluids can sustain counter-rotation globally so that the condition 
$\Omega_{\rm n} / \Omega_{\rm p} < 0$ can be achieved in neutron stars. Nevertheless, the~breaking of the Q-Love relation in these extreme situations is interesting theoretically as it is still not 
properly understood why various universal relations can exist at all, though~some suggestions 
have been proposed~\cite{Yagi:2014a,Sham:2015}. Any examples of the breaking of universal relations might provide hints on the origin of the universality. 
It has been found that the I-Love-Q relations become less accurate if the 
ellipticity of isodensity surfaces of a neutron star displays a large variation inside the star
~\cite{Yagi:2014a}, such as the case for a hot protoneutron star~\cite{Martinon:2014}. } 
In Figure~\ref{fig:epsilon}, we show the profiles of ellipticity $e$ \cite{Hartle:1968}, normalized by 
the dimensionless spin parameter $a$, of~constant energy-density surfaces for four stellar models constructed from the NL3 EOS with $\ln {\bar \lambda}_{\rm tid} = 5$ and different values of $\Omega_{\rm n}/\Omega_{\rm p}$. 
It is seen that, similar to the case of $\Omega_{\rm n}/\Omega_{\rm p}=1$, the~profiles for $\Omega_{\rm n}/\Omega_{\rm p} = -1$ and 10 are nearly constant over a large part of the star. As~we have seen in Figure~\ref{fig:QLove_extreme}, the~Q-Love data for these cases are still relatively insensitive to the EOS models, though~the data for $\Omega_{\rm n}/\Omega_{\rm p}=-1$ deviate largely from the single-fluid Q-Love relation.  
On the other hand, the~case of $\Omega_{\rm n}/\Omega_{\rm p}=0.1$ has a more significant variation in the ellipticity profile, which correlates to the observation that the Q-Love data in this case are more sensitive to the EOS models as shown in Figure~\ref{fig:QLove_extreme}.  
Our results are thus consistent with the suggestion that the breaking of the I-Love-Q relations is
correlated with a large variation of the ellipticity~\cite{Yagi:2014a}.

\begin{figure}[t]
 {\includegraphics[width=340pt]{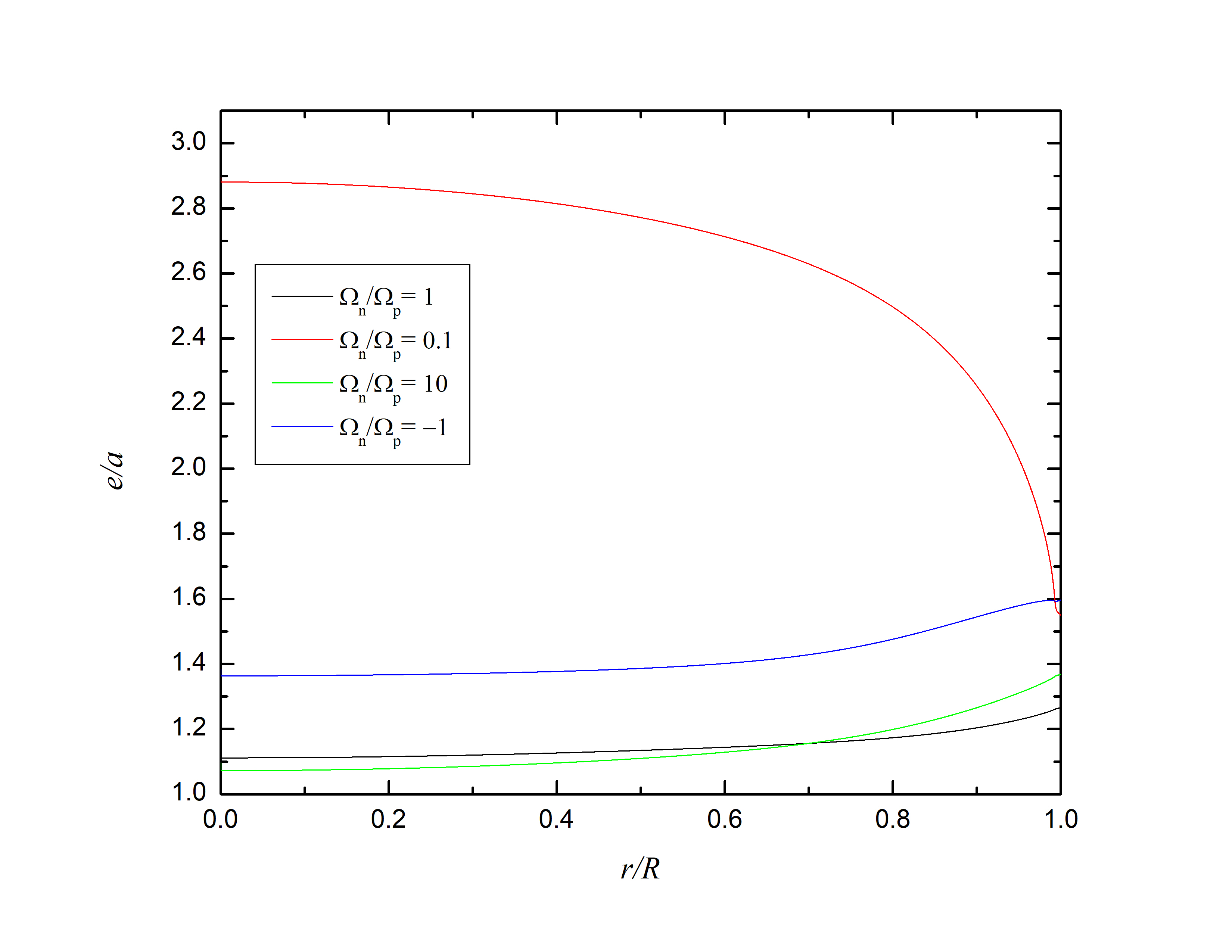}}
  \caption{ Profiles of the ellipticity $e$ (normalized by the spin parameter $a$)
  of constant energy-density surfaces for stellar models constructed from the NL3 EOS with
  $\ln {\bar \lambda}_{\rm tid} = 5$ and different values of $\Omega_{\rm n}/\Omega_{\rm p}$. 
   }
  \label{fig:epsilon}
\end{figure}
\unskip

%%%%%%%%%%%%%%%%%%%%%%%%%%%%%%%%%%%%%%%%%%
\section{Discussion}
\label{sec:discuss}

In summary, we have studied the I-Love-Q relations for superfluid neutron stars based on a general
relativistic two-fluid formulation. 
When neutrons become superfluid, they decouple from the charged components to 
a first approximation. We model the interior of a superfluid neutron star by a two-fluid system containing a neutron superfluid and a ''proton'' fluid containing all charged particles. The~
neutron and proton fluids are also assumed to be uniformly rotating with angular velocities $\Omega_{\rm n}$ and $\Omega_{\rm p}$, respectively. 
The spin ratio $\Omega_{\rm n} / \Omega_{\rm p}$ is the most important parameter in this work.   
We see from the results of our chosen EOS models that the I-Love-Q relations originally 
discovered for single-fluid ordinary neutron stars~\cite{Yagi:2013a,Yagi:2013b}
are still satisfied to high accuracy for superfluid neutron stars when the two fluids are nearly 
co-rotating so that $\Omega_{\rm n} /\Omega_{\rm p} \approx 1$. However, it is also seen that the errors between our superfluid data and the universal relations increase as the spin ratio deviates from unity in both directions (i.e., $\Omega_{\rm n}/\Omega_{\rm p} > 1$ or $<1$). 
If $\Omega_{\rm n} / \Omega_{\rm p}$ could be different from unity by a few tens of percent, the~deviation of the Q-Love relation can be as large as $O(10\%)$ and this may have implications 
for gravitational waveform models that make use of this relation (as discussed in Section~\ref{sec:intro}).

How much can the spin ratio $\Omega_{\rm n} /\Omega_{\rm p}$ actually deviate from unity? 
We can try to address this question by making a connection to the pulsar glitch phenomenon. 
The neutron superfluid inside a spinning isolated neutron star is nearly decoupled from the proton fluid, which spins down continuously due to electromagnetic radiation. The~neutron superfluid is 
thus expected to spin faster (i.e., $\Omega_{\rm n} / \Omega_{\rm p} > 1$). 
When the lag $\delta \Omega = \Omega_{\rm n} - \Omega_{\rm p}$ increases to some critical value, the~Magnus force induced on the superfluid vortices becomes strong enough to lead to an unpinning of the vortices. The~neutron superfluid will then spin down and the proton fluid will spin up as a result of conservation of angular momentum, leading to the glitch phenomenon. 
As a first approximation, it seems reasonable to expect that the lag $\delta \Omega$ is 
comparable to the change in the spin frequency during a glitch. The~fractional change of 
spin frequency $\Delta \nu / \nu$ observed in pulsar glitches ranges from $10^{-11}$ to $10^{-5}$~\cite{Haskell:2015}. This would imply that $\Omega_{\rm n} / \Omega_{\rm p} = 1$ is satisfied to high 
accuracy for isolated pulsars and the deviations of the universal relations due to the two-fluid dynamics will not be observed~astrophysically. 

However, as~one of our goals is to study whether the Q-Love relation remain robust enough to be
used in waveform modeling for binary neutron star systems, it is instructive to also discuss 
the situation in binary systems. 
A new physical process that may come into play in binary systems comparing to isolated stars is the possibility of mass transfer between the two stars. While it is not expected that mass transfer can occur when a binary neutron star system enters the LIGO sensitivity band during the early inspiral phase, it is possible that the first-born neutron star of the system might have spun up due to accretion of matter by Roche-lobe overflow when its companion (i.e., the~second-born neutron star) entered the post main-sequence evolution phase. In~this case, the~proton fluid can spin up and rotate faster than the internal neutron superfluid, leading to the situation 
$\Omega_{\rm n} / \Omega_{\rm p} < 1$.
It is unclear whether this reverse condition, comparing to $\Omega_{\rm n} / \Omega_{\rm p} > 1$, could potentially increase the deviation of the spin ratio away from unity. 
However, there may already be some hints from~observations. 

Recently, three anti-glitches (i.e., $\Delta \nu < 0$) in the accreting pulsar NGC 300 ULX-1 have been observed by {\it NICER} \cite{Ray:2019} and their magnitudes 
$|\Delta \nu / \nu | \sim 10^{-4}$ are significantly larger than typical glitches in isolated radio pulsars. Similarly, a~glitch of magnitude $|\Delta \nu/\nu | \sim 10^{-3}$ has also been seen in the accreting pulsar SXP 1062~\cite{Serim:2017}. 
The normal component is expected to spin faster than the neutron superfluid due to accretion in 
these systems and hence $\Omega_{\rm n} / \Omega_{\rm p} < 1$. 
If the glitches in these accreting pulsars are caused by the transfer of angular momentum between the
superfluid and normal components of the star, with~the assumption that the lag $\delta \Omega$ between the two fluids is also characterized by the glitch magnitude, then these recent 
observations would suggest that the lag magnitude $|\delta \Omega |$ that can be sustained 
between the two fluids could be much larger than that suggested by the glitches in isolated~pulsars. 

While it may be too optimistic (if not unrealistic) to expect that the spin ratio 
$\Omega_{\rm n} /\Omega_{\rm p}$ for neutron stars in binary systems could differ greatly from 
unity, to~the level that the deviation of the Q-Love relation can be as large as $O(10\%)$, a~more 
detailed analysis of this issue would require one to study not only the coupling between the normal and superfluid components at the mesoscopic level inside a neutron star, but~also the formation and evolution of binary neutron star systems—issues that are currently far from completely~understood.

%%%%%%%%%%%%%%%%%%%%%%%%%%%%%%%%%%%%%%%%%%
\vspace{6pt}

%%%%%%%%%%%%%%%%%%%%%%%%%%%%%%%%%%%%%%%%%%
\authorcontributions{
Methodology and formal analysis, C.H.Y.; supervision, L.M.L.; draft preparation, C.H.Y., L.M.L.; review and editing, C.H.Y., L.M.L., N.A., G.C. 
All authors have read and agreed to the published version of the~manuscript.}%For research articles with several authors, a short paragraph specifying their individual contributions must be provided. The following statements should be used ``Conceptualization, X.X. and Y.Y.; methodology, X.X.; software, X.X.; validation, X.X., Y.Y. and Z.Z.; formal analysis, X.X.; investigation, X.X.; resources, X.X.; data curation, X.X.; writing---original draft preparation, X.X.; writing---review and editing, X.X.; visualization, X.X.; supervision, X.X.; project administration, X.X.; funding acquisition, Y.Y. All authors have read and agreed to the published version of the manuscript.'', please turn to the  \href{http://img.mdpi.org/data/contributor-role-instruction.pdf}{CRediT taxonomy} for the term explanation. Authorship must be limited to those who have contributed substantially to the work~reported.
%
%%%%%%%%%%%%%%%%%%%%%%%%%%%%%%%%%%%%%%%%%%%
\funding{This work is partially supported by a grant from the Research Grant Council of the 
Hong Kong Special Administrative Region, China (Project No. 14300320). N.A. acknowledges funding 
by STFC through grant number ST/R00045X/1.}
\conflictsofinterest{The authors declare no conflict of~interest.} 

%%%%%%%%%%%%%%%%%%%%%%%%%%%%%%%%%%%%%%%%%%
%% optional
%\abbreviations{The following abbreviations are used in this manuscript:\\

%\noindent 
%\begin{tabular}{@{}ll}
%MDPI & Multidisciplinary Digital Publishing Institute\\
%DOAJ & Directory of open access journals\\
%TLA & Three letter acronym\\
%LD & linear dichroism
%\end{tabular}}

%%%%%%%%%%%%%%%%%%%%%%%%%%%%%%%%%%%%%%%%%%
%% optional
%\appendixtitles{no} % Leave argument "no" if all appendix headings stay EMPTY (then no dot is printed after "Appendix A"). If the appendix sections contain a heading then change the argument to "yes".
%\appendix
%\section{}
%\unskip
%\subsection{}

%%%%%%%%%%%%%%%%%%%%%%%%%%%%%%%%%%%%%%%%%%
\end{paracol}
\reftitle{References}

%%%%%%%%%%%%%%%%%%%%%%%%%%%%%%%%%%%%%%%%%%

%\externalbibliography{yes}
%\bibliography{sample.bib}

\begin{thebibliography}{999}
\providecommand{\natexlab}[1]{#1}

\bibitem[{LIGO Scientific Collaboration} \em{et~al.}(2017){LIGO Scientific
Collaboration}, { Virgo~Collaboration}, Abbott, Abbott, Abbott, Acernese,
Ackley, Adams, Adams, Addesso, Adhikari, Adya, and et~al]{GW170817}
{LIGO Scientific Collaboration}; { Virgo~Collaboration}; Abbott, B.P.;
Abbott, R.; Abbott, T.D.; Acernese, F.; Ackley, K.; Adams, C.; Adams, T.;
Addesso, P.; et~al.
\newblock GW170817: Observation of Gravitational Waves from a Binary Neutron
Star Inspiral.
\newblock {\em Phys. Rev. Lett.} {\bf 2017}, {\em 119},~161101. [\href{http://doi.org/10.1103/PhysRevLett.119.161101}{CrossRef}]

\bibitem[Annala \em{et~al.}(2018)Annala, Gorda, Kurkela, and
Vuorinen]{Annala:2018}
Annala, E.; Gorda, T.; Kurkela, A.; Vuorinen, A.
\newblock Gravitational-Wave Constraints on the Neutron-Star-Matter Equation of
State.
\newblock {\em Phys. Rev. Lett.} {\bf 2018}, {\em 120},~172703. [\href{http://dx.doi.org/10.1103/PhysRevLett.120.172703}{CrossRef}] %[\href{http://www.ncbi.nlm.nih.gov/pubmed/29756823}{PubMed}]

\bibitem[De \em{et~al.}(2018)De, Finstad, Lattimer, Brown, Berger, and
Biwer]{De:2018}
De, S.; Finstad, D.; Lattimer, J.M.; Brown, D.A.; Berger, E.; Biwer, C.M.
\newblock Tidal Deformabilities and Radii of Neutron Stars from the Observation
of GW170817.
\newblock {\em Phys. Rev. Lett.} {\bf 2018}, {\em 121},~091102. [\href{http://dx.doi.org/10.1103/PhysRevLett.121.091102}{CrossRef}]

\bibitem[Fattoyev \em{et~al.}(2018)Fattoyev, Piekarewicz, and
Horowitz]{Fattoyev:2018}
Fattoyev, F.J.; Piekarewicz, J.; Horowitz, C.J.
\newblock Neutron Skins and Neutron Stars in the Multimessenger Era.
\newblock {\em Phys. Rev. Lett.} {\bf 2018}, {\em 120},~172702. [\href{http://dx.doi.org/10.1103/PhysRevLett.120.172702}{CrossRef}] %[\href{http://www.ncbi.nlm.nih.gov/pubmed/29756822}{PubMed}]

\bibitem[Most \em{et~al.}(2018)Most, Weih, Rezzolla, and
Schaffner-Bielich]{Most:2018}
Most, E.R.; Weih, L.R.; Rezzolla, L.; Schaffner-Bielich, J.
\newblock New Constraints on Radii and Tidal Deformabilities of Neutron Stars
from GW170817.
\newblock {\em Phys. Rev. Lett.} {\bf 2018}, {\em 120},~261103. [\href{http://dx.doi.org/10.1103/PhysRevLett.120.261103}{CrossRef}] %[\href{http://www.ncbi.nlm.nih.gov/pubmed/30004744}{PubMed}]

\bibitem[Tews \em{et~al.}(2018)Tews, Margueron, and Reddy]{Tews:2018}
Tews, I.; Margueron, J.; Reddy, S.
\newblock Critical examination of constraints on the equation of state of dense
matter obtained from GW170817.
\newblock {\em Phys. Rev. C} {\bf 2018}, {\em 98},~045804. [\href{http://dx.doi.org/10.1103/PhysRevC.98.045804}{CrossRef}]

\bibitem[Lim and Holt(2018)]{Lim:2018}
Lim, Y.; Holt, J.W.
\newblock Neutron Star Tidal Deformabilities Constrained by Nuclear Theory and
Experiment.
\newblock {\em Phys. Rev. Lett.} {\bf 2018}, {\em 121},~062701. [\href{http://dx.doi.org/10.1103/PhysRevLett.121.062701}{CrossRef}] %[\href{http://www.ncbi.nlm.nih.gov/pubmed/30141641}{PubMed}]

\bibitem[Malik \em{et~al.}(2018)Malik, Alam, Fortin, Provid\^encia, Agrawal,
Jha, Kumar, and Patra]{Malik:2018}
Malik, T.; Alam, N.; Fortin, M.; Provid\^encia, C.; Agrawal, B.K.; Jha, T.K.;
Kumar, B.; Patra, S.K.
\newblock GW170817: Constraining the nuclear matter equation of state from the
neutron star tidal deformability.
\newblock {\em Phys. Rev. C} {\bf 2018}, {\em 98},~035804. [\href{http://dx.doi.org/10.1103/PhysRevC.98.035804}{CrossRef}]

\bibitem[Li \em{et~al.}(2019)Li, Krastev, Wen, and Zhang]{Li:2019}
Li, B.A.; Krastev, P.G.; Wen, D.H.; Zhang, N.B.
\newblock Towards understanding astrophysical effects of nuclear symmetry
energy.
\newblock {\em Eur. Phys. J. A} {\bf 2019}, {\em 55},~117. [\href{http://dx.doi.org/10.1140/epja/i2019-12780-8}{CrossRef}]

\bibitem[Carson \em{et~al.}(2019)Carson, Steiner, and Yagi]{Carson:2019}
Carson, Z.; Steiner, A.W.; Yagi, K.
\newblock Constraining nuclear matter parameters with GW170817.
\newblock {\em Phys. Rev. D} {\bf 2019}, {\em 99},~043010. [\href{http://dx.doi.org/10.1103/PhysRevD.99.043010}{CrossRef}]

\bibitem[Tsui and Leung(2005)]{Tsui:2005}
Tsui, L.K.; Leung, P.T.
\newblock Probing the Interior of Neutron Stars with Gravitational Waves.
\newblock {\em Phys. Rev. Lett.} {\bf 2005}, {\em 95},~151101. [\href{http://dx.doi.org/10.1103/PhysRevLett.95.151101}{CrossRef}]

\bibitem[Lau \em{et~al.}(2010)Lau, Leung, and Lin]{Lau:2010}
Lau, H.K.; Leung, P.T.; Lin, L.M.
\newblock Inferring physical parameters of compact stars from their f-mode
gravitational wave signals.
\newblock {\em Astrophys. J.} {\bf 2010}, {\em 714},~1234. [\href{http://dx.doi.org/10.1088/0004-637X/714/2/1234}{CrossRef}]

\bibitem[Yagi and Yunes(2013{\natexlab{a}})]{Yagi:2013a}
Yagi, K.; Yunes, N.
\newblock I-Love-Q: Unexpected Universal Relations for Neutron Stars and Quark
Stars.
\newblock {\em Science} {\bf 2013}, {\em 341},~365--368. [\href{http://dx.doi.org/10.1126/science.1236462}{CrossRef}]

\bibitem[Yagi and Yunes(2013{\natexlab{b}})]{Yagi:2013b}
Yagi, K.; Yunes, N.
\newblock I-Love-Q relations in neutron stars and their applications to
astrophysics, gravitational waves, and fundamental physics.
\newblock {\em Phys. Rev. D} {\bf 2013}, {\em 88},~023009. [\href{http://dx.doi.org/10.1103/PhysRevD.88.023009}{CrossRef}]

\bibitem[Yagi(2014)]{Yagi:2014}
Yagi, K.
\newblock Multipole Love relations.
\newblock {\em Phys. Rev. D} {\bf 2014}, {\em 89},~043011. [\href{http://dx.doi.org/10.1103/PhysRevD.89.043011}{CrossRef}]

\bibitem[Chan \em{et~al.}(2014)Chan, Sham, Leung, and Lin]{Chan:2014}
Chan, T.K.; Sham, Y.H.; Leung, P.T.; Lin, L.M.
\newblock Multipolar universal relations between $f$-mode frequency and tidal
deformability of compact stars.
\newblock {\em Phys. Rev. D} {\bf 2014}, {\em 90},~124023. [\href{http://dx.doi.org/10.1103/PhysRevD.90.124023}{CrossRef}]

\bibitem[Chakrabarti \em{et~al.}(2014)Chakrabarti, Delsate, G\``urlebeck, and
Steinhoff]{Chakrabarti:2014}
Chakrabarti, S.; Delsate, T.; G\``urlebeck, N.; Steinhoff, J.
\newblock $I\text{\ensuremath{-}}Q$ Relation for Rapidly Rotating Neutron
Stars.
\newblock {\em Phys. Rev. Lett.} {\bf 2014}, {\em 112},~201102. [\href{http://dx.doi.org/10.1103/PhysRevLett.112.201102}{CrossRef}]

\bibitem[Pappas and Apostolatos(2014)]{Pappas:2014}
Pappas, G.; Apostolatos, T.A.
\newblock Effectively Universal Behavior of Rotating Neutron Stars in General
Relativity Makes Them Even Simpler than Their Newtonian Counterparts.
\newblock {\em Phys. Rev. Lett.} {\bf 2014}, {\em 112},~121101. [\href{http://dx.doi.org/10.1103/PhysRevLett.112.121101}{CrossRef}]

\bibitem[Pappas(2015)]{Pappas:2015}
Pappas, G.
\newblock {Unified description of astrophysical properties of neutron stars
independent of the equation of state}.
\newblock {\em Mon. Not. R. Astron. Soc.} {\bf 2015}, {\em 454},~4066--4084. [\href{http://dx.doi.org/10.1093/mnras/stv2218}{CrossRef}]

\bibitem[Breu and Rezzolla(2016)]{Breu:2016}
Breu, C.; Rezzolla, L.
\newblock {Maximum mass, moment of inertia and compactness of relativistic
stars}.
\newblock {\em Mon. Not. R. Astron. Soc.} {\bf 2016}, {\em 459},~646--656. [\href{http://dx.doi.org/10.1093/mnras/stw575}{CrossRef}]

\bibitem[Bozzola \em{et~al.}(2017)Bozzola, Stergioulas, and
Bauswein]{Bozzola:2017}
Bozzola, G.; Stergioulas, N.; Bauswein, A.
\newblock {Universal relations for differentially rotating relativistic stars
at the threshold to collapse}.
\newblock {\em Mon. Not. R. Astron. Soc.} {\bf 2017}, {\em 474},~3557--3564. [\href{http://dx.doi.org/10.1093/mnras/stx3002}{CrossRef}]

\bibitem[Luk and Lin(2018)]{Luk:2018}
Luk, S.S.; Lin, L.M.
\newblock Universal Relations for Innermost Stable Circular Orbits around
Rapidly Rotating Neutron Stars.
\newblock {\em Astrophys. J.} {\bf 2018}, {\em 861},~141. [\href{http://dx.doi.org/10.3847/1538-4357/aac8d6}{CrossRef}]

\bibitem[Riahi \em{et~al.}(2019)Riahi, Kalantari, and Rueda]{Riahi:2019}
Riahi, R.; Kalantari, S.Z.; Rueda, J.A.
\newblock Universal relations for the Keplerian sequence of rotating neutron
stars.
\newblock {\em Phys. Rev. D} {\bf 2019}, {\em 99},~043004. [\href{http://dx.doi.org/10.1103/PhysRevD.99.043004}{CrossRef}]

\bibitem[Sun \em{et~al.}(2020)Sun, Wen, and Wang]{Sun:2020}
Sun, W.; Wen, D.; Wang, J.
\newblock New quasiuniversal relations for static and rapid rotating neutron
stars.
\newblock {\em Phys. Rev. D} {\bf 2020}, {\em 102},~023039. [\href{http://dx.doi.org/10.1103/PhysRevD.102.023039}{CrossRef}]

\bibitem[Yagi and Yunes(2017)]{Yagi:2017}
Yagi, K.; Yunes, N.
\newblock Approximate universal relations for neutron stars and quark stars.
\newblock {\em Phys. Rep.} {\bf 2017}, {\em 681},~1. [\href{http://dx.doi.org/10.1016/j.physrep.2017.03.002}{CrossRef}]

\bibitem[Doneva and Pappas(2018)]{Doneva:2018}
Doneva, D.D.; Pappas, G. Universal Relations and Alternative Gravity Theories.
\newblock In {\em The Physics and Astrophysics of Neutron Stars}; Rezzolla, L.,
Pizzochero, P., Jones, D.I., Rea, N., Vida{\~{n}}a, I., Eds.; Springer
International Publishing: Cham, Switzerland, 2018; pp.~737--806. [\href{http://dx.doi.org/10.1007/978-3-319-97616-7_13}{CrossRef}]

\bibitem[Lackey \em{et~al.}(2019)Lackey, P\``urrer, Taracchini, and
Marsat]{Lackey:2019}
Lackey, B.D.; P\"{u}rrer, M.; Taracchini, A.; Marsat, S.
\newblock Surrogate model for an aligned-spin effective-one-body waveform model
of binary neutron star inspirals using Gaussian process regression.
\newblock {\em Phys. Rev. D} {\bf 2019}, {\em 100},~024002. [\href{http://dx.doi.org/10.1103/PhysRevD.100.024002}{CrossRef}]

\bibitem[Schmidt and Hinderer(2019)]{Schmidt:2019}
Schmidt, P.; Hinderer, T.
\newblock Frequency domain model of $f$-mode dynamic tides in gravitational
waveforms from compact binary inspirals.
\newblock {\em Phys. Rev. D} {\bf 2019}, {\em 100},~021501. [\href{http://dx.doi.org/10.1103/PhysRevD.100.021501}{CrossRef}]

\bibitem[{Andersson} and {Pnigouras}(2019)]{Andersson:2019}
{Andersson}, N.; {Pnigouras}, P.
\newblock {The seismology of Love: An effective model for the neutron star
tidal deformability}.
\newblock {\em arXiv} {\bf 2019}, arXiv:1905.00012.


\bibitem[Barkett \em{et~al.}(2020)Barkett, Chen, Scheel, and
Varma]{Barkett:2020}
Barkett, K.; Chen, Y.; Scheel, M.A.; Varma, V.
\newblock Gravitational waveforms of binary neutron star inspirals using
post-Newtonian tidal splicing.
\newblock {\em Phys. Rev. D} {\bf 2020}, {\em 102},~024031. [\href{http://dx.doi.org/10.1103/PhysRevD.102.024031}{CrossRef}]

\bibitem[Doneva \em{et~al.}(2013)Doneva, Yazadjiev, Stergioulas, and
Kokkotas]{Doneva:2013}
Doneva, D.D.; Yazadjiev, S.S.; Stergioulas, N.; Kokkotas, K.D.
\newblock Breakdown of I-Love-Q Universality in Rapidly Rotating Relativistic
Stars.
\newblock {\em Astrophys. J.} {\bf 2013}, {\em 781},~L6. [\href{http://dx.doi.org/10.1088/2041-8205/781/1/L6}{CrossRef}]

\bibitem[Haskell \em{et~al.}(2013)Haskell, Ciolfi, Pannarale, and
Rezzolla]{Haskell:2013}
Haskell, B.; Ciolfi, R.; Pannarale, F.; Rezzolla, L.
\newblock On the universality of I-Love-Q relations in magnetized neutron
stars.
\newblock {\em Mon. Not. R. Astron. Soc.} {\bf 2013}, {\em 438},~L71–L75. [\href{http://dx.doi.org/10.1093/mnrasl/slt161}{CrossRef}]

\bibitem[Martinon \em{et~al.}(2014)Martinon, Maselli, Gualtieri, and
Ferrari]{Martinon:2014}
Martinon, G.; Maselli, A.; Gualtieri, L.; Ferrari, V.
\newblock Rotating protoneutron stars: Spin evolution, maximum mass, and
I-Love-Q relations.
\newblock {\em Phys. Rev. D} {\bf 2014}, {\em 90},~064026. [\href{http://dx.doi.org/10.1103/PhysRevD.90.064026}{CrossRef}]

\bibitem[Marques \em{et~al.}(2017)Marques, Oertel, Hempel, and
Novak]{Marques:2017}
Marques, M.; Oertel, M.; Hempel, M.; Novak, J.
\newblock New temperature dependent hyperonic equation of state: Application to
rotating neutron star models and $I\text{\ensuremath{-}}Q$ relations.
\newblock {\em Phys. Rev. C} {\bf 2017}, {\em 96},~045806. [\href{http://dx.doi.org/10.1103/PhysRevC.96.045806}{CrossRef}]

\bibitem[Paschalidis \em{et~al.}(2018)Paschalidis, Yagi, Alvarez-Castillo,
Blaschke, and Sedrakian]{Paschalidis:2018}
Paschalidis, V.; Yagi, K.; Alvarez-Castillo, D.; Blaschke, D.B.; Sedrakian, A.
\newblock Implications from GW170817 and I-Love-Q relations for relativistic
hybrid stars.
\newblock {\em Phys. Rev. D} {\bf 2018}, {\em 97},~084038. [\href{http://dx.doi.org/10.1103/PhysRevD.97.084038}{CrossRef}]

\bibitem[Lau \em{et~al.}(2017)Lau, Leung, and Lin]{Lau:2017}
Lau, S.Y.; Leung, P.T.; Lin, L.M.
\newblock Tidal deformations of compact stars with crystalline quark matter.
\newblock {\em Phys. Rev. D} {\bf 2017}, {\em 95},~101302. [\href{http://dx.doi.org/10.1103/PhysRevD.95.101302}{CrossRef}]

\bibitem[Lau \em{et~al.}(2019)Lau, Leung, and Lin]{Lau:2019}
Lau, S.Y.; Leung, P.T.; Lin, L.M.
\newblock Two-layer compact stars with crystalline quark matter: Screening
effect on the tidal deformability.
\newblock {\em Phys. Rev. D} {\bf 2019}, {\em 99},~023018. [\href{http://dx.doi.org/10.1103/PhysRevD.99.023018}{CrossRef}]

\bibitem[Lombardo and Schulze(2001)]{Lombardo:2001}
Lombardo, U.; Schulze, H.J. Superfluidity in Neutron Star Matter.
\newblock In {\em Physics of Neutron Star Interiors}; Blaschke, D.,
Glendenning, N., Sedrakian, A., Eds.; Springer: Berlin, Germany, 2001; pp.
30--53. [\href{http://dx.doi.org/10.1007/3-540-44578-1_2}{CrossRef}]

\bibitem[Haskell and Sedrakian(2018)]{Haskell:2018}
Haskell, B.; Sedrakian, A. Superfluidity and Superconductivity in Neutron
Stars.
\newblock In {\em The Physics and Astrophysics of Neutron Stars}; Rezzolla, L.,
Pizzochero, P., Jones, D.I., Rea, N., Vida{\~{n}}a, I., Eds.; Springer
International Publishing: Cham, Switzerland, 2018; pp.~401--454. [\href{http://dx.doi.org/10.1007/978-3-319-97616-7_8}{CrossRef}]

\bibitem[Sedrakian and Clark(2019)]{Sedrakian:2019}
Sedrakian, A.; Clark, J.W.
\newblock Superfluidity in nuclear systems and neutron stars.
\newblock {\em Eur. Phys. J. A} {\bf 2019}, {\em 55}, 1--56. [\href{http://dx.doi.org/10.1140/epja/i2019-12863-6}{CrossRef}]

\bibitem[{Haskell} and {Melatos}(2015)]{Haskell:2015}
{Haskell}, B.; {Melatos}, A.
\newblock {Models of pulsar glitches}.
\newblock {\em Int. J. Mod. Phys. D} {\bf 2015}, {\em 24},~1530008. [\href{http://dx.doi.org/10.1142/S0218271815300086}{CrossRef}]

\bibitem[{Anderson} and {Itoh}(1975)]{Anderson:1975}
{Anderson}, P.W.; {Itoh}, N.
\newblock {Pulsar glitches and restlessness as a hard superfluidity
phenomenon}. {\em Nature} {\bf 1975}, {\em 256},~25--27. [\href{http://dx.doi.org/10.1038/256025a0}{CrossRef}]

\bibitem[{Baym} \em{et~al.}(1969){Baym}, {Pethick}, and {Pines}]{Baym:1969}
{Baym}, G.; {Pethick}, C.; {Pines}, D.
\newblock {Superfluidity in Neutron Stars}.
\newblock {\em Nature} {\bf 1969}, {\em 224},~673--674. [\href{http://dx.doi.org/10.1038/224673a0}{CrossRef}]

\bibitem[Yu and Weinberg(2016)]{Yu:2016}
Yu, H.; Weinberg, N.N.
\newblock {Resonant tidal excitation of superfluid neutron stars in coalescing
binaries}.
\newblock {\em Mon. Not. R. Astron. Soc.} {\bf 2016},
{\em 464},~2622--2637. [\href{http://dx.doi.org/10.1093/mnras/stw2552}{CrossRef}]

\bibitem[Sekiguchi \em{et~al.}(2011)Sekiguchi, Kiuchi, Kyutoku, and
Shibata]{Sekiguchi:2011}
Sekiguchi, Y.; Kiuchi, K.; Kyutoku, K.; Shibata, M.
\newblock Effects of Hyperons in Binary Neutron Star Mergers.
\newblock {\em Phys. Rev. Lett.} {\bf 2011}, {\em 107},~211101. [\href{http://dx.doi.org/10.1103/PhysRevLett.107.211101}{CrossRef}]

\bibitem[Bernuzzi \em{et~al.}(2016)Bernuzzi, Radice, Ott, Roberts, M\``osta, and
Galeazzi]{Bernuzzi:2016}
Bernuzzi, S.; Radice, D.; Ott, C.D.; Roberts, L.F.; M\``osta, P.; Galeazzi, F.
\newblock How loud are neutron star mergers?
\newblock {\em Phys. Rev. D} {\bf 2016}, {\em 94},~024023. [\href{http://dx.doi.org/10.1103/PhysRevD.94.024023}{CrossRef}]

\bibitem[Perego \em{et~al.}(2019)Perego, Bernuzzi, and Radice]{Perego:2019}
Perego, A.; Bernuzzi, S.; Radice, D.
\newblock Thermodynamics conditions of matter in neutron star mergers.
\newblock {\em  Eur. Phys. J. A} {\bf 2019}, {\em 55}. [\href{http://dx.doi.org/10.1140/epja/i2019-12810-7}{CrossRef}]

\bibitem[Andersson and Comer(2020)]{Andersson:2020}
Andersson, N.; Comer, G.L.
\newblock Relativistic fluid dynamics: Physics for many different scales. \emph{arXiv}
\textbf{2020}, arXiv:gr-qc/2008.12069.

\bibitem[Gittins \em{et~al.}(2020)Gittins, Andersson, and
Pereira]{Gittins:2020}
Gittins, F.; Andersson, N.; Pereira, J.P.
\newblock Tidal deformations of neutron stars with elastic crusts.
\newblock {\em Phys. Rev. D} {\bf 2020}, {\em 101},~103025. [\href{http://dx.doi.org/10.1103/PhysRevD.101.103025}{CrossRef}]

\bibitem[Carter(1989)]{Carter:1989}
Carter, B.
\newblock Covariant theory of conductivity in ideal fluid or solid media.
\newblock  In \emph{Relativistic Fluid Dynamics}; Anile, A.M., Choquet-Bruhat, Y., Eds.;
Springer: Berlin/Heidelberg, Germany, 1989; pp.~1--64. [\href{http://dx.doi.org/10.1007/BFb0084028}{CrossRef}]

\bibitem[Comer and Langlois(1993)]{Comer:1993}
Comer, G.L.; Langlois, D.
\newblock Hamiltonian formulation for multi-constituent relativistic perfect
fluids.
\newblock {\em Class. Quantum Gravity} {\bf 1993}, {\em 10},~2317--2327. [\href{http://dx.doi.org/10.1088/0264-9381/10/11/014}{CrossRef}]

\bibitem[Comer and Langlois(1994)]{Comer:1994}
Comer, G.L.; Langlois, D.
\newblock Hamiltonian formulation for relativistic superfluids.
\newblock {\em Class. Quantum Gravity} {\bf 1994}, {\em 11},~709--721. [\href{http://dx.doi.org/10.1088/0264-9381/11/3/021}{CrossRef}]

\bibitem[Carter and Langlois(1998)]{Carter:1998}
Carter, B.; Langlois, D.
\newblock Relativistic models for superconducting-superfluid mixtures.
\newblock {\em Nucl. Phys. B} {\bf 1998}, {\em 531},~478--504. [\href{http://dx.doi.org/10.1016/S0550-3213(98)00430-1}{CrossRef}]

\bibitem[Langlois \em{et~al.}(1998)Langlois, Sedrakian, and
Carter]{Langlois:1998}
Langlois, D.; Sedrakian, D.M.; Carter, B.
\newblock {Differential rotation of relativistic superfluid in neutron stars}.
\newblock {\em Mon. Not. R. Astron. Soc.} {\bf 1998}, {\em 297},~1189--1201. [\href{http://dx.doi.org/10.1046/j.1365-8711.1998.01575.x}{CrossRef}]

\bibitem[Andersson(2021)]{Andersson:2021}
Andersson, N.
\newblock A Superfluid Perspective on Neutron Star Dynamics.
\newblock {\em Universe} {\bf 2021}, {\em 7}, 17. [\href{http://dx.doi.org/10.3390/universe7010017}{CrossRef}]

\bibitem[Comer \em{et~al.}(1999)Comer, Langlois, and Lin]{Comer:1999}
Comer, G.L.; Langlois, D.; Lin, L.M.
\newblock Quasinormal modes of general relativistic superfluid neutron stars.
\newblock {\em Phys. Rev. D} {\bf 1999}, {\em 60},~104025. [\href{http://dx.doi.org/10.1103/PhysRevD.60.104025}{CrossRef}]

\bibitem[Andersson and Comer(2001)]{Andersson:2001}
Andersson, N.; Comer, G.L.
\newblock Slowly rotating general relativistic superfluid neutron stars.
\newblock {\em Class. Quantum Gravity} {\bf 2001}, {\em 18},~969--1002. [\href{http://dx.doi.org/10.1088/0264-9381/18/6/302}{CrossRef}]

\bibitem[Char and Datta(2018)]{Char:2018}
Char, P.; Datta, S.
\newblock Relativistic tidal properties of superfluid neutron stars. {\em Phys. Rev. D} {\bf 2018}, {\em 98}, 084010. 084010. [\href{http://dx.doi.org/10.1103/PhysRevD.98.084010}{CrossRef}]

\bibitem[Andersson \em{et~al.}(2002)Andersson, Comer, and
Langlois]{Andersson:2002}
Andersson, N.; Comer, G.L.; Langlois, D.
\newblock Oscillations of general relativistic superfluid neutron stars.
\newblock {\em Phys. Rev. D} {\bf 2002}, {\em 66},~104002. [\href{http://dx.doi.org/10.1103/PhysRevD.66.104002}{CrossRef}]

\bibitem[Lin \em{et~al.}(2008)Lin, Andersson, and Comer]{Lin:2008}
Lin, L.M.; Andersson, N.; Comer, G.L.
\newblock Oscillations of general relativistic multifluid/multilayer compact
stars.
\newblock {\em Phys. Rev. D} {\bf 2008}, {\em 78},~083008. [\href{http://dx.doi.org/10.1103/PhysRevD.78.083008}{CrossRef}]

\bibitem[Prix \em{et~al.}(2005)Prix, Novak, and Comer]{Prix:2005}
Prix, R.; Novak, J.; Comer, G.L.
\newblock Relativistic numerical models for stationary superfluid neutron
stars.
\newblock {\em Phys. Rev. D} {\bf 2005}, {\em 71},~043005. [\href{http://dx.doi.org/10.1103/PhysRevD.71.043005}{CrossRef}]

\bibitem[Sourie \em{et~al.}(2016)Sourie, Oertel, and Novak]{Sourie:2016}
Sourie, A.; Oertel, M.; Novak, J.
\newblock Numerical models for stationary superfluid neutron stars in general
relativity with realistic equations of state.
\newblock {\em Phys. Rev. D} {\bf 2016}, {\em 93},~083004. [\href{http://dx.doi.org/10.1103/PhysRevD.93.083004}{CrossRef}]

\bibitem[{Hartle}(1967)]{Hartle:1967}
{Hartle}, J.B.
\newblock {Slowly Rotating Relativistic Stars. I. Equations of Structure}.
\newblock {\em Astrophys. J.} {\bf 1967}, {\em 150},~1005. [\href{http://dx.doi.org/10.1086/149400}{CrossRef}]


\bibitem[Hinderer(2008)]{Hinderer:2008}
Hinderer, T.
\newblock Tidal Love Numbers of Neutron Stars.
\newblock {\em Astrophys. J.} {\bf 2008}, {\em 677},~1216--1220. [\href{http://dx.doi.org/10.1086/533487}{CrossRef}]

\bibitem[Damour and Nagar(2009)]{Damour:2009}
Damour, T.; Nagar, A.
\newblock Relativistic tidal properties of neutron stars.
\newblock {\em Phys. Rev. D} {\bf 2009}, {\em 80},~084035. [\href{http://dx.doi.org/10.1103/PhysRevD.80.084035}{CrossRef}]

\bibitem[Pani \em{et~al.}(2015)Pani, Gualtieri, and Ferrari]{Pani:2015}
Pani, P.; Gualtieri, L.; Ferrari, V.
\newblock Tidal Love numbers of a slowly spinning neutron star.
\newblock {\em Phys. Rev. D} {\bf 2015}, {\em 92},~124003. [\href{http://dx.doi.org/10.1103/PhysRevD.92.124003}{CrossRef}]

\bibitem[Landry(2017)]{Landry:2017}
Landry, P.
\newblock Tidal deformation of a slowly rotating material body: Interior metric
and Love numbers.
\newblock {\em Phys. Rev. D} {\bf 2017}, {\em 95},~124058. [\href{http://dx.doi.org/10.1103/PhysRevD.95.124058}{CrossRef}]


\bibitem[Datta and Char(2020)]{Datta:2020}
Datta, S.; Char, P.
\newblock Effect of superfluid matter of a neutron star on the tidal
deformability.
\newblock {\em Phys. Rev. D} {\bf 2020}, {\em 101},~064016. [\href{http://dx.doi.org/10.1103/PhysRevD.101.064016}{CrossRef}]

\bibitem[Comer and Joynt(2003)]{Comer:2003}
Comer, G.L.; Joynt, R.
\newblock Relativistic mean field model for entrainment in general relativistic
superfluid neutron stars.
\newblock {\em Phys. Rev. D} {\bf 2003}, {\em 68},~023002. [\href{http://dx.doi.org/10.1103/PhysRevD.68.023002}{CrossRef}]

\bibitem[Comer(2004)]{Comer:2004}
Comer, G.L.
\newblock Slowly rotating general relativistic superfluid neutron stars with
relativistic entrainment.
\newblock {\em Phys. Rev. D} {\bf 2004}, {\em 69},~123009. [\href{http://dx.doi.org/10.1103/PhysRevD.69.123009}{CrossRef}]

\bibitem[Kheto and Bandyopadhyay(2014)]{Kheto:2014}
Kheto, A.; Bandyopadhyay, D.
\newblock Isospin dependence of entrainment in superfluid neutron stars in a
relativistic model.
\newblock {\em Phys. Rev. D} {\bf 2014}, {\em 89},~023007. [\href{http://dx.doi.org/10.1103/PhysRevD.89.023007}{CrossRef}]

\bibitem[Fattoyev \em{et~al.}(2010)Fattoyev, Horowitz, Piekarewicz, and
Shen]{Fattoyev:2010}
Fattoyev, F.J.; Horowitz, C.J.; Piekarewicz, J.; Shen, G.
\newblock Relativistic effective interaction for nuclei, giant resonances, and
neutron stars.
\newblock {\em Phys. Rev. C} {\bf 2010}, {\em 82},~055803. [\href{http://dx.doi.org/10.1103/PhysRevC.82.055803}{CrossRef}]

\bibitem[Glendenning and Moszkowski(1991)]{Glendenning:1991}
Glendenning, N.K.; Moszkowski, S.A.
\newblock Reconciliation of neutron-star masses and binding of the
\ensuremath{\Lambda} in hypernuclei.
\newblock {\em Phys. Rev. Lett.} {\bf 1991}, {\em 67},~2414--2417. [\href{http://dx.doi.org/10.1103/PhysRevLett.67.2414}{CrossRef}]

\bibitem[Yagi \em{et~al.}(2014)Yagi, Stein, Pappas, Yunes, and
Apostolatos]{Yagi:2014a}
Yagi, K.; Stein, L.C.; Pappas, G.; Yunes, N.; Apostolatos, T.A.
\newblock Why I-Love-Q: Explaining why universality emerges in compact objects.
\newblock {\em Phys. Rev. D} {\bf 2014}, {\em 90},~063010. [\href{http://dx.doi.org/10.1103/PhysRevD.90.063010}{CrossRef}]

\bibitem[Sham \em{et~al.}(2015)Sham, Chan, Lin, and Leung]{Sham:2015}
Sham, Y.H.; Chan, T.K.; Lin, L.M.; Leung, P.T.
\newblock Unveiling the university of I-Love-Q relations.
\newblock {\em Astrophys. J.} {\bf 2015}, {\em 798},~121. [\href{http://dx.doi.org/10.1088/0004-637X/798/2/121}{CrossRef}]

\bibitem[{Hartle} and {Thorne}(1968)]{Hartle:1968}
{Hartle}, J.B.; {Thorne}, K.S.
\newblock {Slowly Rotating Relativistic Stars. II. Models for Neutron Stars and
Supermassive Stars}.
\newblock {\em Astrophys. J.} {\bf 1968}, {\em 153},~807. [\href{http://dx.doi.org/10.1086/149707}{CrossRef}]

\bibitem[Ray \em{et~al.}(2019)Ray, Guillot, Ho, Kerr, Enoto, Gendreau,
Arzoumanian, Altamirano, Bogdanov, Campion, Chakrabarty, Deneva, Jaisawal,
Kozon, Malacaria, Strohmayer, and Wolff]{Ray:2019}
Ray, P.S.; Guillot, S.; Ho, W.C.G.; Kerr, M.; Enoto, T.; Gendreau, K.C.;
Arzoumanian, Z.; Altamirano, D.; Bogdanov, S.; Campion, R.; et al.
\newblock Anti-glitches in the Ultraluminous Accreting Pulsar {NGC} 300 {ULX}-1
Observed with {NICER}.
\newblock {\em Astrophys. J.} {\bf 2019}, {\em 879},~130. [\href{http://dx.doi.org/10.3847/1538-4357/ab24d8}{CrossRef}]

\bibitem[Serim \em{et~al.}(2017)Serim, Şahiner, Çerri Serim, İnam, and
Baykal]{Serim:2017}
Serim, M.M.; Şahiner, Ş.; Çerri-Serim, D.; İnam, S.Ç.; Baykal, A.L.
\newblock {Discovery of a glitch in the accretion-powered pulsar SXP 1062}.
\newblock {\em Mon. Not. R. Astron. Soc.} {\bf 2017}, {\em 471},~4982--4989. [\href{http://dx.doi.org/10.1093/mnras/stx1771}{CrossRef}]


\end{thebibliography}

%=====================================
% References, variant B: internal bibliography
%=====================================
%\begin{thebibliography}{999}

% Reference 1
%\bibitem[Author1(year)]{ref-journal}
%Author1, T. The title of the cited article. {\em Journal Abbreviation} {\bf 2008}, {\em 10}, 142--149.
% Reference 2
%\bibitem[Author2(year)]{ref-book}
%Author2, L. The title of the cited contribution. In {\em The Book Title}; Editor1, F., Editor2, A., Eds.; Publishing House: City, Country, 2007; pp. 32--58.

%\end{thebibliography}

%%%%%%%%%%%%%%%%%%%%%%%%%%%%%%%%%%%%%%%%%%
\end{document}